\documentclass[reprint,aps,prb,superscriptaddress,nobibnotes]{revtex4-1}
\usepackage{amsmath}
\usepackage{amssymb}
\usepackage{natbib}
\usepackage{graphicx}
\usepackage{color}
\usepackage{colordvi}
\usepackage{bm}
\usepackage{calrsfs}
\usepackage[thinlines,thiklines]{easybmat}

\setcitestyle{numbers,square}

\begin{document}

\title{Surface dynamics of rough magnetic films} 
\author{Tao Yu} 
\affiliation{Kavli Institute of NanoScience, Delft University of Technology, 2628 CJ Delft, The Netherlands} 
\author{Sanchar Sharma} 
\affiliation{Kavli Institute of NanoScience, Delft University of Technology, 	2628 CJ Delft, The Netherlands} 
\author{Yaroslav M. Blanter} 
\affiliation{Kavli Institute of NanoScience, Delft University of Technology, 	2628 CJ Delft, The Netherlands} 
\author{Gerrit E. W. Bauer} 
\affiliation{Institute for Materials Research $\&$ WPI-AIMR $\&$ CSRN, Tohoku University, Sendai 980-8577, Japan}   
\affiliation{Kavli Institute of 	NanoScience, Delft University of Technology, 2628 CJ Delft, The Netherlands} \date{\today }  

\begin{abstract}
The chirality of magnetostatic Damon-Eshbach (DE) magnons affects the transport of energy and angular momentum at the surface of magnetic films and spheres. We calculate the surface-disorder-limited dephasing and transport lifetimes of surface modes of sufficiently thick high-quality ferromagnetic films such as that of yttrium iron garnet. Surface magnons are not protected by chirality, but interact strongly with smooth surface roughness. Nevertheless, for long-range disorder, the transport is much less affected by the suppressed back scattering (vertex correction). Moreover, in the presence of roughness, ferromagnetic resonance under a \textit{uniform} microwave field can gennerate a considerable amount of surface magnons.
\end{abstract}

 \maketitle

%\thanks{Author to whom correspondence should be addressed}  

\section{Introduction}  

Spin waves or its quanta, the magnons, are weakly dissipating carriers of angular momentum and energy \cite{magnonics1,magnonics2,magnonics3,magnonics4,spin_caloritronics}. The magnetostatic surface or Damon-Eshbach (DE) spin waves in finite-size magnets have additional unique features \cite{Walker,DE,spin_waves_book,new_book} such as exponential localization at the surface of spheres \cite{Walker} or films \cite{DE} and directional chirality: the surface magnons propagate only in one direction that is governed by surface normal and magnetization directions \cite{Walker,DE,spin_waves_book,new_book}. The surface magnons are, for example, found to transport heat in particular directions, even against a temperature gradient, i.e., heat conveyer-belt effect \cite{heatconveyer1,heatconveyer2,heatconveyer3,heatconveyer4}. In spherical magnetic resonators surface magnons can strongly interact with optical whispering gallery modes \cite{WGM1,WGM2,WGM3,Sanchar_PRB}, where the chirality of the DE mode can be beneficial for magnon cooling by light \cite{optical_cooling}.  

The physics of surface magnons depends on their lifetime and mean-free path that are limited by disorder, phonon and magnon scattering  \cite{magnonics1,magnonics2,magnonics3,magnonics4} and especially by surface roughness \cite{static1,static2,static3,static4}. On the other hand, although surface magnons are formally not topologically protected, the effect of chirality on the magnon lifetime is an interesting issue. Previous studies \cite{static1,static3,static4} focused on the damping of the uniform spin precession (the Kittel mode) by two-magnon scattering at disorder in either bulk materials \cite{static1} or films with nearly zero thickness \cite{static3,static4}. Recently, scattering of the dipole-exchange spin waves by single edge defects in very thin films ($80$~nm) was studied by numerically solving the linearized Landau-Lifshitz equations, reporting a suppression of backscattering of chiral spin waves in the DE configuration, even though the magnetization amplitude is nearly constant over the film \cite{backscattering_immune}.  

Here we quasi-analytically study lifetime and transport of chiral DE magnons, i.e., for a configuration in which the spin waves propagate normal to an in-plane magnetization, in the presence of surface disorder. We focus on magnetic films/slabs which are so thick that surface states are well established, but thin enough to allow interactions between surfaces. We find that surface roughness strongly reduces the lifetime of magnons, but does not affect the transport mildly because of suppressed backscattering. Furthermore, we propose that surface roughness allows for an efficient population of surface states by a uniform microwave field. An asymmetry of the surface roughness on the two surfaces of the film \cite{static4,Tao_MoS2} can lead to an unbalanced excitation
 of the surface magnons on opposite sides of the sample, which is a necessary conditions for the magnon conveyer-belt \cite{heatconveyer1,heatconveyer2,heatconveyer3,heatconveyer4}.  

This paper is organized as follows. We first review the equations that govern the magnon amplitudes or wave functions in Sec.~\ref{wavefunction}. In Sec.~\ref{interaction}, we establish scattering cross sections for the magnon--surface-roughness interaction via the dipolar and Zeeman interactions. The magnon and spin transport lifetimes of surface magnons are addressed in Secs.~\ref{damping}, \ref{transport} and \ref{excitation}. We summarize the results and give an outlook in Sec.~\ref{summary}.  

\section{Surface magnon wave function} \label{wavefunction}

Magnetostatic waves in ferromagnetic films with in-plane magnetization were studied long back \cite{DE,spin_waves_book}. Here, we review the amplitude or \textquotedblleft wave function\textquotedblright\ of the surface magnons as far as it is relevant for our objectives. As shown in Fig. \ref{surface_roughness}, the surface is perpendicular to $\hat{\mathbf{x}}$-axis, the equilibrium magnetization points along $\hat{\mathbf{z}}$-axis, and we are mainly interested in spin waves propagating along $ \hat{\mathbf{y}}$-axis.  

The magnetization $\mathbf{M}(\mathbf{r})$ satisfies the Landau-Lifshitz (LL) equation \cite{Landau}  
\begin{equation}
	{\partial\mathbf{M}(\mathbf{r})}/{\partial t} =  - \gamma\mu_0\mathbf{M}(\mathbf{r})\times\mathbf{H}_{\mathrm{eff}}(\mathbf{r}), \label{Heisenberg} 
\end{equation} 
where $\gamma$ is modulus of the gyromagnetic ratio, $\mu_0$ denotes the vacuum permeability, and the effective magnetic field,  
\begin{equation}
	 \mathbf{H}_{\mathrm{eff}}(\mathbf{r}) =  - (1/\mu_0){\delta F}\left[ \mathbf{M} \right] /{\delta\mathbf{M}}(\mathbf{r}), 
\end{equation} 
with $F$ being the free energy functional. In the presence of an applied magnetic field $H_z\hat{\mathbf{z}}$ and dipolar interactions,  
\begin{equation}
	 F =  -{\mu_0}\int d\mathbf{r}\left[ M_zH_z + \frac{\mathbf{M}(\mathbf{r})}{8\pi}\cdot\nabla\int d\mathbf{r}^{\prime}\frac{\nabla^{\prime}\cdot\mathbf{M} (\mathbf{r}^{\prime})}{|\mathbf{r} - \mathbf{r}^{\prime}|}\right] . \label{free_energy} 
\end{equation} 
We disregard the crystalline anisotropy and damping, which is often allowed in high quality materials such as yttrium iron garnet (YIG) \cite{static4,magnetization1,magnetization2}. We ignore exchange interaction which is valid for the spin waves with wavelengths much larger than exchange wavelength $\sim 100 $nm in YIG \cite{exchange_stiffness}. 

We linearize Eq.~(\ref{Heisenberg}) for small magnetization amplitudes around $\mathbf{M} = M_0\hat{\mathbf{z}},$ where $M_0$ is the saturation magnetization. For a film with in-plane translation symmetry $\mathbf{M}_{\gamma = x,y}^{j\mathbf{k}} = m_{\gamma}^{j\mathbf{k}}(x)e^{ik_yy}e^{ik_zz} $, where  
\begin{align}
	 m_x^{j\mathbf{k}}(x) &= a_{j\mathbf{k}}e^{i\kappa_jx} + b_{j\mathbf{k}}e^{- i\kappa_jx}, \\
	 m_y^{j\mathbf{k}}(x) &= c_{j\mathbf{k}}e^{i\kappa_jx} + d_{j\mathbf{k}}e^{- i\kappa_jx} . 
\end{align} 
Here, $j$ labels the energy bands of the magnons and $\mathbf{k} = k_y\hat{\mathbf{y}} + k_z\hat{\mathbf{z}}$ represents the in-plane momentum. We choose the normalization condition \cite{Walker,magnetic_nanodots}  
\begin{equation}
	 \int d\mathbf{r}[M_x^{j\mathbf{k}}(\mathbf{r})M_y^{j\mathbf{k}}(\mathbf{r})^{\ast} - M_x^{j\mathbf{k}}(\mathbf{r})^{\ast}M_y^{j\mathbf{k}}(\mathbf{r})] =  - i/2, \label{normalization2} 
\end{equation} 
in which $M^{\ast}$ is the complex conjugate of $M$. The coefficients $ \left\{a,b,c,d\right\}_{j\mathbf{k}}$ are determined with the ansatz $ \mathbf{M}_{\gamma}^{j\mathbf{k}}\propto e^{- i\omega_{j\mathbf{k}}t}$ in which $\omega_{j\mathbf{k}}$ is the eigen-frequency. The linearized LL equations are,
\begin{align}
\hspace{-0.3cm}\left(
\begin{array}{cc}
i\omega_{j\mathbf{k}}- \omega_M \frac{\kappa_{j}k_{y}}{k_{s}^{2}} & 
-\omega_H- \omega_M \frac{k_{y}^{2}}{k_{s}^{2}} \\ 
\omega_H + \omega_M \frac{\kappa_{j}^{2}}{k_{s}^{2}} & 
i\omega_{j\mathbf{k}} + \omega_M \frac{\kappa_{j}k_{y}}{k_{s}^{2}}%
\end{array}
\right) \left(
\begin{array}{c}
a_{j\mathbf{k}} \\ 
c_{j\mathbf{k}}%
\end{array}
\right) & =0, \label{abRes} \\
%\hspace{-0.3cm}\left(
%\begin{array}{cc}
%i\omega_{j\mathbf{k}}+\mu_{0}\gamma M_{0}\frac{\kappa_{j}k_{y}}{k_{s}^{2}} & 
%-\mu_{0}\gamma H_{z}-\mu_{0}\gamma M_{0}\frac{k_{y}^{2}}{k_{s}^{2}} \\ 
%\mu_{0}\gamma H_{z}+\mu_{0}\gamma M_{0}\frac{\kappa_{j}^{2}}{k_{s}^{2}} & 
%i\omega_{j\mathbf{k}}-\mu_{0}\gamma M_{0}\frac{\kappa_{j}k_{y}}{k_{s}^{2}}%
%\end{array}
%\right) \left(
%\begin{array}{c}
%b_{j\mathbf{k}} \\ 
%d_{j\mathbf{k}}%
%\end{array}
%\right) & =0,  \notag \\
\left(
\begin{array}{cc}
f_{+}(\mathbf{k}) & f_{+}^{\ast}(\mathbf{k}) \\ 
f_{-}^{\ast}(\mathbf{k}) & f_{-}(\mathbf{k})%
\end{array}
\right) \left(
\begin{array}{c}
a_{j\mathbf{k}} \\ 
b_{j\mathbf{k}}%
\end{array}
\right) & =0,   \label{AC}
\end{align}
where $\omega_H = \gamma\mu_0 H_z$, $\omega_M = \gamma \mu_0 M_0$,
\begin{equation}
	 f_{\pm}(\mathbf{k}) = \frac{1}{2}\left(|\mathbf{k}|\pm ik_y\frac{i\omega_{j \mathbf{k}}/\mu_0 - \gamma M_0\kappa_jk_y/k_s^2}{\gamma H_z + \gamma M_0k_y^2/k_s^2}\right) \frac{e^{i\kappa_j{d}/{2}}}{i\kappa_j - |\mathbf{k}|} 
\end{equation} 
and $k_s^2 = \kappa_j^2 + |\mathbf{k}|^2$. A similar equation as (\ref{abRes}) holds with $\{a_{j\mathbf{k}},c_{j\mathbf{k}},\kappa_j\} \rightarrow \{b_{j\mathbf{k}},d_{j\mathbf{k}},-\kappa_j\}$. Eq. (\ref{abRes}) gives the dispersion relation \cite{DE}  
\begin{equation}
	 \omega_{j\mathbf{k}} = \sqrt{\omega_H^2 + \omega_H \omega_M \frac{\kappa_j^2 + k_y^2}{k_s^2} }. \label{energy_spectra} 
\end{equation} 
Eq.~(\ref{AC}) gives the characteristic equation for $\kappa_j$ \cite{DE},  
\begin{equation}
	 (\beta k_y)^2 + \kappa_j^2(\alpha + 1)^2 - |\mathbf{k}|^2 - 2\kappa_j| \mathbf{k}|(\alpha + 1)\cot(\kappa_jd) = 0 . \label{characteristic} 
\end{equation} 
Here, $\alpha = \omega_H \omega_M/(\omega_H^2 - \omega_{j\mathbf{k}}^2)$ and $\beta = \omega_{j\mathbf{k}} \omega_M/(\omega_H^2 - \omega^2)$.  

When $\kappa_j = iq_x$ is purely imaginary, we obtain a surface \textquotedblleft DE\textquotedblright\ mode \cite{DE}:  
\begin{align}
	 m_x^{\mathbf{k}}(x) &= C\left[ e^{- q_xx}(- \alpha q_x + \beta k_y) + De^{q_xx}(\alpha q_x + \beta k_y)\right] , \notag \\
	 m_y^{\mathbf{k}}(x) &= iC\left[ e^{- q_xx}(- \beta q_x + \alpha k_y) + De^{q_xx}(\beta q_x + \alpha k_y)\right] , \label{wavefunctions_DE} 
\end{align} 
in which $C$ is governed by the normalization Eq.~(\ref{normalization2}); $d$ denotes the thickness of the film; and  
\begin{equation}
	 D = e^{- q_xd}\frac{q_x(\alpha + 1) - \beta k_y - \left\vert \mathbf{k} \right\vert}{q_x(\alpha + 1) + \beta k_y + \left\vert \mathbf{k}\right\vert} . 
\end{equation} 
The characteristic relation for the DE mode becomes \cite{DE},  
\begin{equation}
	 (\beta k_y)^2 - q_x^2(\alpha + 1)^2 - \left\vert \mathbf{k}\right\vert^2 - 2q_x|\mathbf{k}|(\alpha + 1)\coth(q_xd) = 0 . \label{ce_de} 
\end{equation}

For surface magnons with momenta $k_y\hat{\mathbf{y}} + k_z\hat{\mathbf{z}} =  - |k_y|\hat{\mathbf{y}} + k_z\hat{\mathbf{z}}$ for $q_xd\gtrsim1$:  
\begin{equation}
	 im_x^{\mathbf{k}} + \frac{k_y}{\left\vert \mathbf{k}\right\vert} m_y^{\mathbf{k}} \approx 0, \label{chiral_relation} 
\end{equation} 
i.e., when $\mathbf{k} = k_y\hat{\mathbf{y}}$ the DE magnons are circularly polarized. When the wave vector has a small component along the saturated magnetization, i.e., $\mathbf{k} = k_y\hat{\mathbf{y}} + \delta k_z\hat{\mathbf{z}}$, the DE modes precess elliptically. From Eq.~(\ref{ce_de}) we conclude that DE modes preserve their character as long as $\left\vert \delta k_z\right\vert <\left\vert k_y\right\vert \sqrt{M_0/H_z}$  \cite{DE}. We now prove that for small $\delta k_z$ the ellipticity is weak: When $q_xd\gtrsim1$, $\coth (q_xd)\rightarrow1$ and Eq.~(\ref {ce_de}) simplifies to  
\begin{equation}
	 q_x(\alpha + 1) + \left\vert \mathbf{k}\right\vert \approx |\beta k_y| = \beta k_y, 
\end{equation} 
which implies $D\rightarrow0$ in Eq.~(\ref{wavefunctions_DE}). Therefore  
\begin{align}
	 & im_x^{\mathbf{k}} + \frac{k_y}{\left\vert \mathbf{k}\right\vert} m_y^{\mathbf{k}} \notag \\
	 & \rightarrow iCe^{- q_xx} \left[ - \alpha q_x + \beta k_y + \frac{k_y}{|\mathbf{k} |}(- \beta q_x + \alpha k_y)\right] \notag \\ &= \frac{iCe^{- q_xx}}{\left\vert \mathbf{k}\right\vert } [\left\vert \mathbf{k} \right\vert^2 - (\alpha + 1)q_x^2 + \alpha k_y^2] = 0, \label{chiral2} 
\end{align} 
where the term in square brackets vanishes because of the dispersion relation Eq.~(\ref{energy_spectra}). This relation is essential for the chiral coupling between the magnons and surface roughness as discussed in Sec.~\ref{interaction}.  

\section{Magnon--surface-roughness interaction} \label{interaction}

\begin{figure}[th]
\begin{center}
{\includegraphics[width=8.5cm]{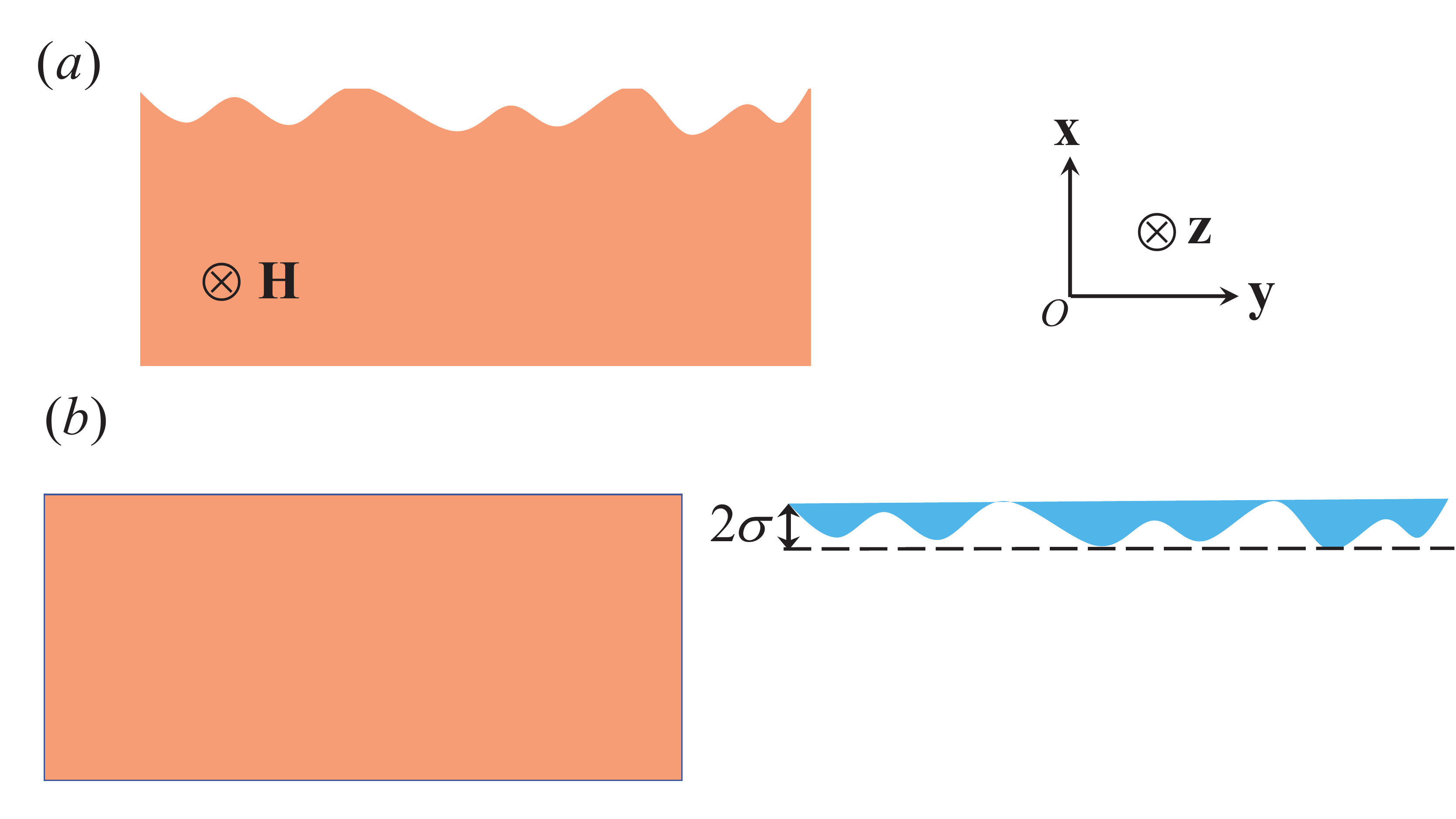}}
\end{center}
\caption{(Color online) (a) Surface roughness on the upper surface of a
magnetic film. The surface normal is along the $\hat{\mathbf{x}}$-direction.
(b) The roughness is located only in a thin surface layer on top of an
ordered magnetic film with thickness that corresponds to twice the
root-mean-square $\protect\sigma$ of the thickness fluctuations.}
\label{surface_roughness}
\end{figure}

While surface disorder has many manifestations, we focus on a simple generic model due to lack of precise information: the magnetic order is preserved up to the surface position that varies slightly as a function of position in a random manner. The film with surface roughness [Fig.~\ref {surface_roughness}(a)] can be separated into two parts: the smooth film and a fluctuating thin surface layer \cite{static1,static2,static3,static4} , as shown in Fig.~\ref{surface_roughness}(b).  

The free energy in Eq.~(\ref{free_energy}) is affected by the surface fluctuation, inducing two magnon scattering. Below, we derive the magnon--surface-roughness interaction due to the dipolar and Zeeman interactions.  

\subsection{Dipolar interaction} \label{dipolar}

The free energy due to the dipolar interaction in the magnetic film \cite{Landau}  
\begin{equation}
	 F_d =  - \frac{\mu_0}{2}\int d\mathbf{r}_{\parallel}\int_{- \frac{d}{2} + x_l(\mathbf{r}_{\parallel})}^{\frac{d}{2} + x_u(\mathbf{r}_{\parallel})}dx \mathbf{M}(\mathbf{r})\cdot\mathbf{H}_D(\mathbf{r}), \label{free_dipolar} 
\end{equation} 
where $\mathbf{r}_{\parallel} = y\hat{\mathbf{y}} + z\hat{\mathbf{z}}$. $x_u(\mathbf{r}_{\parallel})$ and $x_l(\mathbf{r}_{\parallel})$ are the fluctuating part of the the upper and lower surface positions, respectively. The magnetic potential $\psi$, defined in terms of the demagnetization field as $\mathbf{H}_D =  - \nabla\psi$, can be written as Coulomb-like expression \cite{Landau}  
\begin{equation}
	 \psi(\mathbf{r}) =  - \int_Vd\mathbf{r^{\prime}}\frac{\nabla\cdot{\mathbf{M}(\mathbf{r^{\prime}})}}{{4\pi}|{\mathbf{r} - \mathbf{r^{\prime}}}|} . 
\end{equation} 
The free energy reads  
\begin{align}
	 F_d &=  - \frac{\mu_0}{8\pi}\int d\mathbf{r}_{\parallel}\int_{- \frac{d}{2} + x_l(\mathbf{r}_{\parallel})}^{\frac{d}{2} + x_u(\mathbf{r}_{\parallel})}dx\int d\mathbf{r}_{\parallel}^{\prime}\int_{- \frac{d}{2} + x_l(\mathbf{r}_{\parallel}^{\prime})}^{\frac{d}{2} + x_u(\mathbf{r}_{\parallel}^{\prime})}dx^{\prime} \notag \\
	 & \times M_{\beta}(\mathbf{r})\partial_{\beta}\partial_{\alpha}\frac{M{_{\alpha}(\mathbf{r}^{\prime})}}{\left\vert{\mathbf{r} - \mathbf{r}^{\prime}} \right\vert}, \label{free_dipolar2} 
\end{align} 
using the summation convention over repeated Cartesian indices $\alpha = \{x,y,z\}$. When the amplitudes of $x_u(\mathbf{r}_{\parallel})$ and $ x_l(\mathbf{r}_{\parallel})$ are much smaller than both thickness of the film and decay depth of the DE modes, we can simplify Eq.~(\ref {free_dipolar2}) by the \textit{mean-value theorem} for the integral, i.e.,  
\begin{equation}
	 \int_{d/2}^{d/2 + x_u}f(x)dx \approx f(\frac{d}{2})x_u . 
\end{equation} 
To linear order, $F_d = F_0 + F_d^u + F_d^l$, where $F_0$ is given by Eq. (\ref{free_dipolar2}) putting $x_u = x_l = 0$,
\begin{align}
	 F_d^u &=  - \frac{\mu_0}{4\pi}\int d\mathbf{r}_{\parallel}\int_{- \frac{d}{2}}^{\frac{d}{2}}dxM_{\beta}(\mathbf{r})\partial_{\beta}\partial_{\alpha} \int d\mathbf{r}_{\parallel}^{\prime}x_u(\mathbf{r}_{\parallel}^{\prime}) \notag \\
	 & \times\frac{M_{\alpha}({d}/{2},\mathbf{r}_{\parallel}^{\prime})}{\sqrt{(x - d/2)^2 + (\mathbf{r}_{\parallel} - \mathbf{r}_{\parallel}^{\prime})^2}}, \label{free_dipolar_u} 
\end{align} 
and 
\begin{align}
	 F_d^l &= \frac{\mu_0}{4\pi}\int d\mathbf{r}_{\parallel}\int_{- \frac{d}{2}}^{\frac{d}{2}}dxM_{\beta}(\mathbf{r})\partial_{\beta}\partial_{\alpha}\int d\mathbf{r}_{\parallel}^{\prime}x_l(\mathbf{r}_{\parallel}^{\prime}) \notag \\
	 & \times\frac{M_{\alpha}(-{d}/{2},\mathbf{r}_{\parallel}^{\prime})}{\sqrt{(x + d/2)^2 + (\mathbf{r}_{\parallel} - \mathbf{r}_{\parallel}^{\prime})^2}} . \label{free_dipolar_l} 
\end{align} 
Note that this approximation does not take the large-momentum scattering into account that is caused by the derivative of $x_{u/l}(\mathbf{r}_{\parallel})$. Our theory is therefore limited to the smooth surface roughness that governs the Gilbert damping \cite{static4}.  

These expressions can be integrated with the Hamiltonian formulation for the magnetization dynamics \cite{squeezed_magnon,spin_waves_book,Kittel_book,HP,Sanchar_PRB}. Substituting $ \mathbf{M}\rightarrow - \gamma\hbar\hat{\mathbf{S}}$ (and $M_0 = \gamma\hbar S$), the Hamiltonian for the upper surface roughness reads \cite{squeezed_magnon,spin_waves_book,Kittel_book,HP,Sanchar_PRB}  
\begin{align}
	 &H_{\mathrm{d}}^u =  - \frac{\mu_0\gamma^2\hbar^2}{4\pi}\int d\mathbf{r} \int d\mathbf{r}_{\parallel}^{\prime}\left(
\begin{array}
{ccc} \hat{S}_x(\mathbf{r}) & \hat{S}_y(\mathbf{r}) & \hat{S}_z(\mathbf{r}) 
\end{array} \right) \notag \\
	 & \times\mathbf{\hat{G}}\left(x - \frac{d}{2},\mathbf{r}_{\parallel} - \mathbf{r}_{\parallel}^{\prime}\right) \left(\hat{S}_x(\frac{d}{2},\mathbf{r}_{\parallel}^{\prime}),\hat{S}_y(\frac{d}{2},\mathbf{r}_{\parallel}^{\prime}),\hat{S}_z(\frac{d}{2},\mathbf{r}_{\parallel}^{\prime})\right)^T, 
\end{align} 
introducing the Green function tensor \cite{Kalinikos}  
\begin{align}
	 & \mathbf{\hat{G}}\left(x - \frac{d}{2},\mathbf{r}_{\parallel} - \mathbf{r}_{\parallel}^{\prime}\right) \notag \\
	 & \equiv\left(
\begin{array}
{ccc} \partial_x^2 & \partial_x\partial_y & \partial_x\partial_z \\ \partial_y\partial_x & \partial_y^2 & \partial_y\partial_z \\ \partial_z\partial_x & \partial_z\partial_y & \partial_z^2 
\end{array} \right) \frac{x_u(\mathbf{r}_{\parallel}^{\prime})}{\sqrt{(x - \frac{d}{2})^2 + (\mathbf{r}_{\parallel} - \mathbf{r}_{\parallel}^{\prime})^2}} . 
\end{align} 
We focus on the linear regime, thereby disregarding higher-order terms encoding magnon-magnon scattering process that become important for large magnon numbers \cite{Suhl_nonlinear,nonlinear2}. The spin operators may then be expressed in terms of magnon operators $\hat{\alpha}_{j\mathbf{k}}$ \cite{squeezed_magnon,Kittel_book,HP,Sanchar_PRB},  
\begin{align}
	 \hat{S}_{x,y}(\mathbf{r}) &= \sqrt{2S}\sum_{j,\mathbf{k}}[M_{x,y}^{j\mathbf{k}}(\mathbf{r})\hat{\alpha}_{j\mathbf{k}} + M_{x,y}^{j\mathbf{k}}(\mathbf{r})^{\ast}\hat{\alpha}_{j\mathbf{k}}^{\dagger}], \notag \\
	 \hat{S}_z(\mathbf{r}) &=  - S + (\hat{S}_x^2 + \hat{S}_y^2)/(2S) . \label{Bogoliubov} 
\end{align} 
The interaction for the upper surface then reduces to  
\begin{align}
	 &H_{\mathrm{d}}^u = \sum_{j\mathbf{k}}\left(L_{j\mathbf{k}}\hat{\alpha}_{j \mathbf{k}} + \mathrm{h . c .}\right) + \sum_{j\mathbf{k}}\sum_{j^{\prime}\mathbf{k}^{\prime}}\left[ A_{j\mathbf{k},j^{\prime}\mathbf{k}^{\prime}}\hat{\alpha}_{j\mathbf{k}}\hat{\alpha}_{j^{\prime}\mathbf{k}^{\prime}}\right . \notag \\
	 & \left .  + B_{j\mathbf{k},j^{\prime}\mathbf{k}^{\prime}}\hat{\alpha}_{j \mathbf{k}}^{\dagger}\hat{\alpha}_{j^{\prime}\mathbf{k}^{\prime}} + C_{j \mathbf{k},j^{\prime}\mathbf{k}^{\prime}}\hat{\alpha}_{j\mathbf{k}}^{\dagger} \hat{\alpha}_{j^{\prime}\mathbf{k}^{\prime}}^{\dagger} + D_{j\mathbf{k} ,j^{\prime}\mathbf{k}^{\prime}}\hat{\alpha}_{j\mathbf{k}}\hat{\alpha}_{j^{\prime}\mathbf{k}^{\prime}}^{\dagger}\right] . \label{HamRough} 
\end{align} 
The coefficients of the linear term 
\begin{align}
	 L_{j\mathbf{k}} &=  - \frac{\mu_0\gamma M_0}{2}\sqrt{\frac{2\hbar M_0}{\gamma}}x_u(- \mathbf{k})k_z\int dxe^{(x - \frac{d}{2})|\mathbf{k}|} \notag \\
	 & \times\left[ im_x^{j\mathbf{k}}(x) + \frac{{k_y}}{{|\mathbf{k}|}}m_y^{j \mathbf{k}}(x)\right] , \label{linear} 
\end{align} 
nearly vanish for DE modes with momenta $k_y\hat{\mathbf{y}} + k_z\hat{\mathbf{z}} =  - |k_y|\hat{\mathbf{y}} + k_z\hat{\mathbf{z}}$ when $|\mathbf{k} |d\gtrsim1$ because of Eq.~(\ref{chiral_relation}). Therefore, although dipolar interaction is long-range, the surface roughness of the \textit{upper} surface has little effect on the surface magnons propagating near the \textit{lower} surface (and vice versa).  

The linear terms do not conserve spin and therefore exert a torque on the magnetization $\mathbf{M}_0(\mathbf{r}) = M_0\hat{\mathbf{z}}$. When the linear term is eliminated by the transformation $\hat{\alpha}_{j\mathbf{k}}^{\dagger}\rightarrow\hat{\alpha}_{j\mathbf{k}}^{\dagger} -{L_{j\mathbf{k}}}/{\omega_{j\mathbf{k}}}$, Eq.~(\ref{Bogoliubov}) introduces transverse components of the equilibrium magnetization  
\begin{equation}
	 \mathbf{M}_{x,y}^0(\mathbf{r}) = \sqrt{2M_0\gamma\hbar}\sum_{j,\mathbf{k}} \left[ \frac{M_{x,y}^{j\mathbf{k}}(\mathbf{r}){L_{j\mathbf{k}}^{\ast}}}{{\omega_{j\mathbf{k}}}} + \mathrm{h . c .}\right] . 
\end{equation} 
Strong surface disorder therefore affects the equilibrium magnetization and eigenmodes. However, here we focus on weak disorder with $\left\vert \mathbf{M}_{x,y}^0(\mathbf{r})\right\vert \ll M_0,$ where we may disregard the linear term.  

The quadratic terms in $H_{\mathrm{d}}^u$ represent two-magnon scattering by disorder with coefficients  
\begin{align}
	 &A_{j\mathbf{k},j^{\prime}\mathbf{k}^{\prime}} =  - \mu_0\gamma\hbar M_0x_u(- \mathbf{k} - \mathbf{k}^{\prime})\left\{\int dxe^{(x - \frac{d}{2})| \mathbf{k}|} \right . \notag \\
	 & \times \left(
\begin{array}
{cc} m_x^{j\mathbf{k}}(x) & m_y^{j\mathbf{k}}(x) 
\end{array} \right) \left(
\begin{array}
{cc} |\mathbf{k}| & - ik_y \\ - ik_y & - \frac{\mathbf{k}_y^2}{|\mathbf{k}|} 
\end{array} \right) \left(
\begin{array}
{c} m_x^{j^{\prime}\mathbf{k}^{\prime}}(\frac{d}{2}) \\ m_y^{j^{\prime}\mathbf{k}^{\prime}}(\frac{d}{2}) 
\end{array} \right) \notag \\ & + \frac{1}{2}\frac{(k_z + k_z^{\prime})^2}{\left\vert \mathbf{k} + \mathbf{k}^{\prime}\right\vert}\int dx\left[ m_x^{j\mathbf{k}}(x)m_x^{j^{\prime} \mathbf{k}^{\prime}}(x) + m_y^{j\mathbf{k}}(x)m_y^{j^{\prime}\mathbf{k}^{\prime}}(x)\right] \notag \\ & \left . \times e^{(x - \frac{d}{2})|\mathbf{k} + \mathbf{k}^{\prime}|} - 2m_x^{j\mathbf{k}}({d}/{2})m_x^{j^{\prime}\mathbf{k}^{\prime}}({d}/{2})\right\} , \label{quadraticA} 
\end{align} 
and  
\begin{align}
	 &B_{j\mathbf{k},j^{\prime}\mathbf{k}^{\prime}} =  - \mu_0\gamma\hbar M_0x_u(\mathbf{k} - \mathbf{k}^{\prime})\left\{\int dxe^{(x - \frac{d}{2})| \mathbf{k}|} \right . \notag \\
	 & \times \left(
\begin{array}
{cc} m_x^{j\mathbf{k}}(x)^{\ast} & m_y^{j\mathbf{k}}(x)^{\ast} 
\end{array} \right) \left(
\begin{array}
{cc} |\mathbf{k}| & ik_y \\ ik_y & - \frac{k_y^2}{|\mathbf{k}|} 
\end{array} \right) \left(
\begin{array}
{c} m_x^{j^{\prime}\mathbf{k}^{\prime}}(\frac{d}{2}) \\ m_y^{j^{\prime}\mathbf{k}^{\prime}}(\frac{d}{2}) 
\end{array} \right) \notag \\ & + \frac{1}{2}\frac{(k_z - k_z^{\prime})^2}{\left\vert \mathbf{k} - \mathbf{k}^{\prime}\right\vert}\int dx\left[ m_x^{j\mathbf{k}}(x)^{\ast}m_x^{j^{\prime}\mathbf{k}^{\prime}}(x) + m_y^{j\mathbf{k}}(x)^{\ast}m_y^{j^{\prime}\mathbf{k}^{\prime}}(x)\right] \notag \\ & \left . \times e^{(x - \frac{d}{2})|\mathbf{k} - \mathbf{k}^{\prime}|} - 2m_x^{j\mathbf{k}}({d}/{2})^{\ast}m_x^{j^{\prime}\mathbf{k}^{\prime}}({d}/{2})\right\} . \label{quadraticB} 
\end{align} 
$C_{j\mathbf{k},j^{\prime}\mathbf{k}^{\prime}} = A_{j\mathbf{k},j^{\prime} \mathbf{k}^{\prime}}^{\ast}$ and $D_{j\mathbf{k},j^{\prime}\mathbf{k}^{\prime}} = B_{j\mathbf{k},j^{\prime}\mathbf{k}^{\prime}}^{\ast}$ by hermiticity. The first term in Eq.~(\ref{quadraticB}) is the interaction by the non-local dipolar interaction that is inefficient between two DE states with opposite momenta due to (nearly) circular polarization $m_x^{j\mathbf{k}}(x)^{\ast} + i\frac{k_y}{|\mathbf{k}|}m_x^{j\mathbf{k}}(x)^{\ast} \approx 0$ when $\mathbf{k} =  - \left\vert k_y\right\vert \hat{\mathbf{y}} + \delta k_z\hat{\mathbf{z}}\ $and $\left\vert \delta k_z\right\vert <\left\vert k_y\right\vert \sqrt{M_0/H_z},\ $as discussed above, by which the first term in Eq.~(\ref{quadraticB}) vanishes. The second and third terms can be traced to the local part of the dipolar interaction. Reflection of DE magnons to the opposite direction (and surface), is exponentially suppressed because the wave function overlap of modes on different surfaces is small. The large momentum backscattering of DE magnons by surface disorder is therefore suppressed by both chirality and nearly circular polarization.  

\subsection{Zeeman energy}  

The free energy due to the Zeeman interaction is \cite{Landau}  
\begin{equation}
	 F_Z =  - \mu_0\int d\mathbf{r}_{\parallel}\int_{- d/2 + x_l(\mathbf{r}_{\parallel})}^{d/2 + x_u(\mathbf{r}_{\parallel})}dx\mathbf{M}(\mathbf{r})\cdot\mathbf{H}_z, 
\end{equation} 
and the equivalent Hamiltonian reads
\begin{equation}
	 H_Z = \frac{\mu_0\gamma\hbar}{2S}\int d\mathbf{r}_{\parallel} \int_{- d/2 + x_l(\mathbf{r}_{\parallel})}^{d/2 + x_u(\mathbf{r}_{\parallel})} \left[ \hat{S}_x^2(\mathbf{r}) + \hat{S}_y^2(\mathbf{r})\right] H_zdx . 
\end{equation} 
As above, we derive the interaction Hamiltonian with small surface roughness  
\begin{align}
	 H_Z^s &= \frac{\mu_0\gamma^2\hbar^2}{2M_0}H_z\int d\mathbf{r}_{\parallel}\left[ \hat{S}_x^2\left(\frac{d}{2},\mathbf{r}_{\parallel}\right) + \hat{S}_y^2\left(\frac{d}{2},\mathbf{r}_{\parallel}\right) \right] x_u(\mathbf{r}_{\parallel}) \notag \\
	 & - \frac{\mu_0\gamma^2\hbar^2}{2M_0}H_z\int d\mathbf{r}_{\parallel}\left[ \hat{S}_x^2\left(\frac{d}{2},\mathbf{r}_{\parallel}\right) + \hat{S}_y^2\left(\frac{d}{2},\mathbf{r}_{\parallel}\right) \right] x_l(\mathbf{r}_{\parallel}) . 
\end{align} 
By the Bogoliubov transformation Eq.~(\ref{Bogoliubov}), the interaction Hamiltonian by a rough upper surface becomes  
\begin{align}
	 H_Z^u &= \sum_{j\mathbf{k}}\sum_{j^{\prime}\mathbf{k}^{\prime}}\left[ \tilde{A}_{j\mathbf{k},j^{\prime}\mathbf{k}^{\prime}}\hat{\alpha}_{j\mathbf{k}}\hat{\alpha}_{j^{\prime}\mathbf{k}^{\prime}} + \tilde{B}_{j\mathbf{k} ,j^{\prime}\mathbf{k}^{\prime}}\hat{\alpha}_{j\mathbf{k}}^{\dagger}\hat{\alpha}_{j^{\prime}\mathbf{k}^{\prime}}\right . \notag \\
	 & \left .  + \tilde{C}_{j\mathbf{k},j^{\prime}\mathbf{k}^{\prime}}\hat{\alpha}_{j\mathbf{k}}^{\dagger}\hat{\alpha}_{j^{\prime}\mathbf{k}^{\prime}}^{\dagger} + \tilde{D}_{j\mathbf{k},j^{\prime}\mathbf{k}^{\prime}} \hat{\alpha}_{j\mathbf{k}}\hat{\alpha}_{j^{\prime}\mathbf{k}^{\prime}}^{\dagger}\right] , 
\end{align} 
in which  
\begin{align}
	 \tilde{A}_{j\mathbf{k},j^{\prime}\mathbf{k}} &= \mu_0\gamma\hbar H_zx_u(- \mathbf{k} - \mathbf{k}^{\prime})\sum_{\gamma = x,y}m_{\gamma}^{j \mathbf{k}}\left(\frac{d}{2}\right) m_{\gamma}^{j^{\prime}\mathbf{k}^{\prime}}\left(\frac{d}{2}\right) , \notag \\
	 \tilde{B}_{j\mathbf{k},j^{\prime}\mathbf{k}} &= \mu_0\gamma\hbar H_zx_u(\mathbf{k} - \mathbf{k}^{\prime})\sum_{\gamma = x,y}m_{\gamma}^{j \mathbf{k}}\left(\frac{d}{2}\right)^{\ast}m_{\gamma}^{j^{\prime}\mathbf{k}^{\prime}}\left(\frac{d}{2}\right) . \label{Zeeman_quadratic} 
\end{align} 
The fluctuations in the Zeeman energy generated by the surface roughness are efficient only when there is significant wave function overlap between states on the same surfaces.  

\section{Surface damping} \label{damping}

We now use the Hamiltonians derived in the previous section to find the damping of surface magnons by rough surfaces.  

\subsection{Analytical analysis}  

The Green function of a magnon in the $j$-th band with in-plane wave-vector $ \mathbf{k}$ is \cite{Fetter,Mahan,Abrikosov}  
\begin{equation}
	 G_{j\mathbf{k}}(\omega) = \frac{1}{\omega - \omega_{j\mathbf{k}} + i\Gamma_{j \mathbf{k}} - \Sigma_{j\mathbf{k}}(\omega)}, 
\end{equation} 
where $\omega_{j\mathbf{k}}$ is the resonance frequency, $\Sigma_{j\mathbf{k}}(\omega)$ is the self-energy due to surface scattering, $\Gamma_{j\mathbf{k}} = \alpha_0\omega_{j\mathbf{k}}$ is the intrinsic damping in the absence of surface roughness, and $\alpha_0$ is the Gilbert damping constant \cite{Gilbert_damping} of the Kittel mode of a film with smooth surfaces. The imaginary part of $\Sigma$ governs the magnon scattering rate or damping due to the surface roughness  
\begin{equation}
	 \alpha_s(\omega_{j\mathbf{k}})\equiv -{2{\rm Im}\Sigma (\omega_{\mathbf{k}})}/{\omega_{\mathbf{k}}} . \label{surface_damping} 
\end{equation}

\begin{figure}[th]
\begin{center}
{\includegraphics[width=7.5cm]{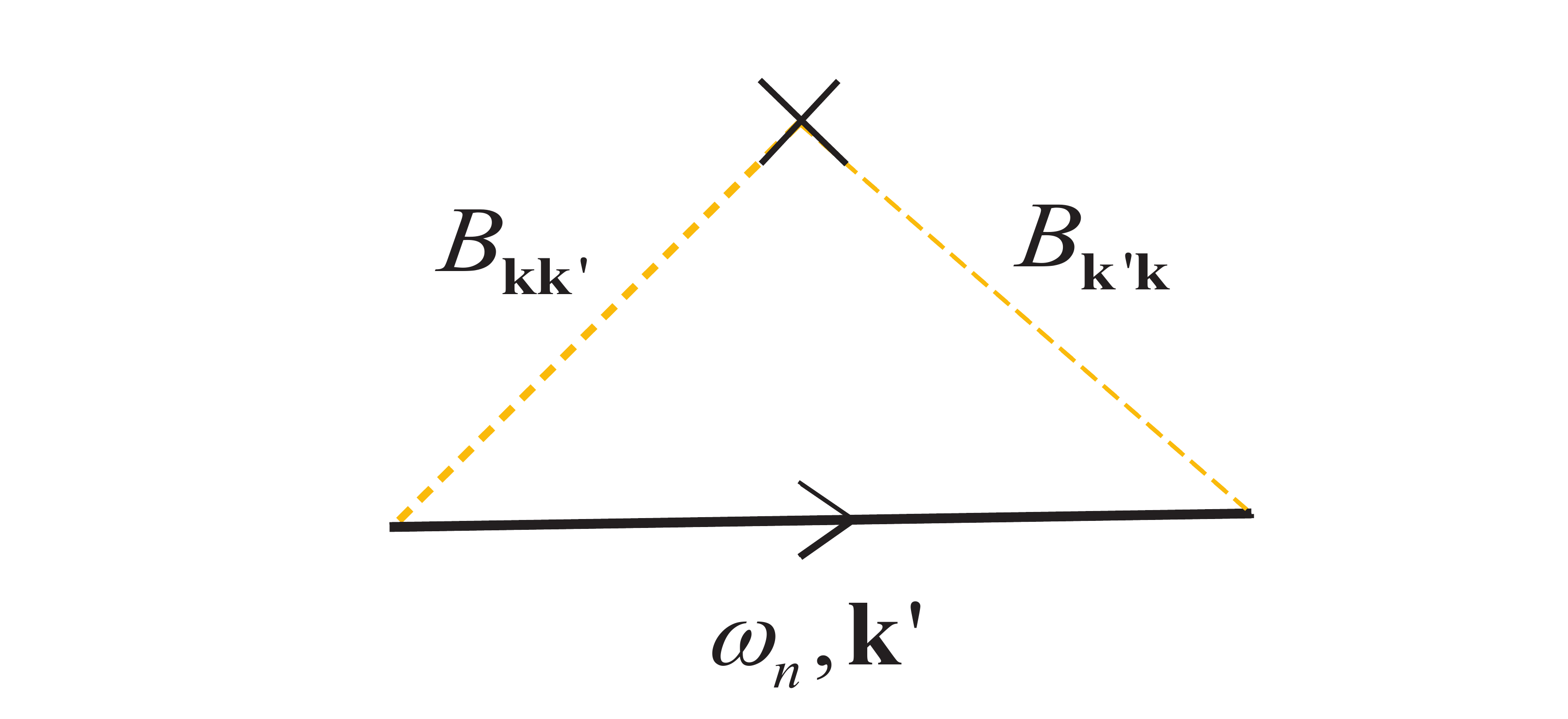}}
\end{center}
\caption{(Color online) Feynman diagram for the self energy in the
self-consistent Born approximation. Here, $\rightarrow$ represents the full
Green function $G_{j\mathbf{k}}(i\protect\omega_{n})$. The orange dashed
line denotes the scattering potential.}
\label{Feynmann}
\end{figure}

In the Matsubara representation \cite{Fetter,Mahan,Abrikosov},  
\begin{align}
	 &G_{j\mathbf{k}}(\tau - \tau^{\prime}) \equiv - \left\langle T_{\tau}\hat{\alpha}_{j\mathbf{k}}(\tau)\hat{\alpha}_{j\mathbf{k}}^{\dagger}(\tau^{\prime})\right\rangle \notag \\
	 &=  - \left\langle T_{\tau}\hat{\alpha}_{j\mathbf{k}}(\tau)\hat{\alpha}_{j \mathbf{k}}^{\dagger}\left(\tau^{\prime}\right) \exp\left(- \int_0^{\beta}d\tilde{\tau}H_{\mathrm{int}}^s\right) \right\rangle , \label{Green_functions} 
\end{align} 
where $\beta = 1/(k_BT)$ and $T$ is the temperature. The first and second lines in Eq.~(\ref{Green_functions}) are expressed in the Heisenberg and interaction representations, respectively. $T_{\tau}$ is the chronological product with imaginary time $\tau$. $H_{\mathrm{int}}^s$ is the interaction Hamiltonian due to surface roughness  
\begin{equation}
	 H_{\mathrm{int}}^s = \sum_{j\mathbf{k}}\sum_{j^{\prime}\mathbf{k}^{\prime}}[ \mathcal{A}_{j\mathbf{k},j^{\prime}\mathbf{k}^{\prime}}\hat{\alpha}_{j \mathbf{k}}\hat{\alpha}_{j^{\prime}\mathbf{k}^{\prime}} + \mathcal{B}_{j \mathbf{k},j^{\prime}\mathbf{k}^{\prime}}\hat{\alpha}_{j\mathbf{k}}^{\dagger} \hat{\alpha}_{j^{\prime}\mathbf{k}^{\prime}}] + \mathrm{h . c .}, \label{TotIntHam}
\end{equation} 
in which $\mathcal{A}_{j\mathbf{k},j^{\prime}\mathbf{k}^{\prime}} = A_{j \mathbf{k},j^{\prime}\mathbf{k}^{\prime}} + \tilde{A}_{j\mathbf{k},j^{\prime} \mathbf{k}^{\prime}}$ and $\mathcal{B}_{j\mathbf{k},j^{\prime}\mathbf{k}^{\prime}} = B_{j\mathbf{k},j^{\prime}\mathbf{k}^{\prime}} + \tilde{B}_{j \mathbf{k},j^{\prime}\mathbf{k}^{\prime}}$. In the \textit{weak} coupling regime, the Green function in the frequency-momentum space $G_{j\mathbf{k}}(i\omega_n) = \int_0^{\beta}d\tau e^{i\omega_n\tau}G_{j\mathbf{k}}(\tau)$ can be expanded in the \textit{self-consistent} Born approximation  \cite{Ando} as  
\begin{align}
	 &G_{j\mathbf{k}}(i\omega_n) = G_{j\mathbf{k}}^{(0)}(i\omega_n) + G_{j \mathbf{k}}^{(0)}(i\omega_n)\left\{\sum_{j^{\prime}\mathbf{k}^{\prime}}| \mathcal{B}_{j^{\prime}\mathbf{k}^{\prime},j\mathbf{k}}|^2\right . \notag \\
	 & \left . \times G_{j^{\prime}\mathbf{k}^{\prime}}(i\omega_n) + \sum_{j^{\prime}\mathbf{k}^{\prime}}|\mathcal{A}_{j^{\prime}\mathbf{k}^{\prime},j\mathbf{k}}|^2G_{j^{\prime}\mathbf{k}^{\prime}}(- i\omega_n) \right\} G_{j\mathbf{k}}(i\omega_n), 
\end{align} 
where $G_{j\mathbf{k}}^{(0)}(i\omega_n) = 1/(i\omega_n - \omega_{j\mathbf{k}} + i\Gamma_{j\mathbf{k}})$. The corresponding Feynman diagrams for the self-energy due to the surface scattering is shown in Fig.~\ref{Feynmann}.  

In the real frequency domain, by the analytical continuation $i\omega_n\rightarrow\omega + i\delta$, the self-energy of the magnons from the surface roughness is calculated to be  
\begin{align}
	 \Sigma_{j\mathbf{k}}(\omega) &= \sum_{j^{\prime}\mathbf{k}^{\prime}}| \mathcal{B}_{j\mathbf{k},j^{\prime}\mathbf{k}^{\prime}}|^2\frac{G_{j^{\prime}\mathbf{k}^{\prime}}^{(0)}(\omega)}{1 - G_{j^{\prime}\mathbf{k}^{\prime}}^{(0)}(\omega)\Sigma_{j^{\prime}\mathbf{k}^{\prime}}(\omega)} \notag \\
	 & + \sum_{j^{\prime}\mathbf{k}^{\prime}}|\mathcal{A}_{j\mathbf{k},j^{\prime} \mathbf{k}^{\prime}}|^2\frac{G_{j^{\prime}\mathbf{k}^{\prime}}^{(0)}(- \omega)}{1 - G_{j^{\prime}\mathbf{k}^{\prime}}^{(0)}(- \omega)\Sigma_{j^{\prime}\mathbf{k}^{\prime}}(- \omega)} . 
\end{align} 
At the magnon's frequency $\omega = \omega_{j\mathbf{k}}$,  
\begin{align}
	 &\Sigma_{j\mathbf{k}}(\omega_{j\mathbf{k}}) = \sum_{j^{\prime}\mathbf{k}^{\prime}}|\mathcal{B}_{j\mathbf{k},j^{\prime}\mathbf{k}^{\prime}}|^2\frac{G_{j^{\prime}\mathbf{k}^{\prime}}^{(0)}(\omega_{j\mathbf{k}})}{1 - G_{j^{\prime}\mathbf{k}^{\prime}}^{(0)}(\omega_{j\mathbf{k}})\Sigma_{j^{\prime}\mathbf{k}^{\prime}}(\omega_{j\mathbf{k}})} \notag \\
	 & + \sum_{j^{\prime}\mathbf{k}^{\prime}}|\mathcal{A}_{j\mathbf{k},j^{\prime} \mathbf{k}^{\prime}}|^2\frac{G_{j^{\prime}\mathbf{k}^{\prime}}^{(0)}(- \omega_{j\mathbf{k}})}{1 - G_{j^{\prime}\mathbf{k}^{\prime}}^{(0)}(- \omega_{j \mathbf{k}})\Sigma_{j^{\prime}\mathbf{k}^{\prime}}(- \omega_{j\mathbf{k}})} . \label{imaginary_self} 
\end{align} 
The $\mathcal{A}$-term is off-resonant, with negligible contribution to the self-energy since $\omega_{j\mathbf{k}} + \omega_{j^{\prime}\mathbf{k}^{\prime}}\gg\Gamma_{j^{\prime}\mathbf{k}^{\prime}}$ in $G_{j^{\prime} \mathbf{k}^{\prime}}^{(0)}(- \omega_{j\mathbf{k}})$. Hence, in the calculation below, we disregard this contribution, which is the \textquotedblleft rotating wave approximation\textquotedblright\ \cite{nonlinear2,input_output1,input_output2}. Using Eq. (\ref{quadraticB}), $| \mathcal{B}_{j\mathbf{k},j^{\prime}\mathbf{k}^{\prime}}|^2\propto x_{u,l}(\mathbf{k} - \mathbf{k}^{\prime})x_{u,l}(\mathbf{k}^{\prime} - \mathbf{k})$ and under the ergodic hypothesis, a configurational averaging of $\Sigma_{j \mathbf{k}}$ over the disorder leads to a self-correlation function that we model by a Gaussian \cite{static4,Tao_MoS2}  
\begin{equation}
	 \left\langle x_{u,l}(\mathbf{k})x_{u,l}(- \mathbf{k})\right\rangle = \pi R_{u,l}^2\sigma_{u,l}^2\exp({- |\mathbf{k}|^2R_{u,l}^2/4}), \label{correlation} 
\end{equation} 
in which $\sigma$ and $R$ are the root-mean-square (rms) of the amplitude and correlation length of the surface roughness, respectively.  

\subsection{Results} \label{numerical_damping}

Concrete predictions for the magnon damping in a specific material require sample and material parameters. We focus on a YIG film with $ \mu_0M_0 = 0 . 177$~T \cite{Haiming_PRL,magnetization1,magnetization2}, $ \alpha_0 = 5\times10^{- 5}$ \cite{YIG_damping1,Haiming_PRL}, and $d = 3$ \textrm{~}$\mathrm{\mu}$m. The surface topology of YIG can be varied by different polishing methods~\cite{surface_sensitivity}. Varying $\sigma$ from several nm and correlation lengths $R$ of the order micrometers, strongly affected the transverse spin Seebeck effect. However, here we focus on longitudinal (in-plane) transport. We adopt here the smooth surface roughness with $R = 2$\thinspace$\mathrm{\mu}$m, $\sigma_u = 4$~nm as reported for ferromagnetic metal films \cite{static4,surface_sensitivity}. The interface to the substrate gallium gadolinium garnet (GGG) is believed to be of very high quality, so we disregard any interface roughness of the lower surface, i.e. adopt $\sigma_l = 0$. The choice for a long-range surface roughness implies that magnetostatic magnons cannot be scattered into (degenerate) exchange-regime magnons with high momentum, which are therefore disregarded in the following.  

The dispersion relations of the surface and bulk modes of magnetic films can be found in Fig.~3 in Ref.~\cite{DE} and many textbooks. DE modes are allowed for finite $\delta k_z$ as long as $\left\vert{\delta k_z}/{k_y}\right\vert <\sqrt{M_0/H_z}$ and frequencies larger than $ \sqrt{\omega_H \omega_M + \omega_H^2}$, \cite{DE}. We focus here on DE magnons with $k_z = 0$ and $|k_y|d\geq1/2$ that are exponentially localized near the surface and frequencies approaching the limiting constant $\omega_H + \omega_M/2$ \cite{spin_waves_book} (see also Sec.~\ref{wavefunction}). These magnons are spectrally distant from the magnetostatic bulk modes with frequency $ \leq \sqrt{\omega_H^2 + \omega_H \omega_M}$ \cite{DE,spin_waves_book}, which therefore do not contribute to the self-energy of the surface magnons by two-magnon scattering, Eq.~(\ref{imaginary_self}).  

\begin{figure}[ht]
\begin{center}
{\includegraphics[width=10.5cm]{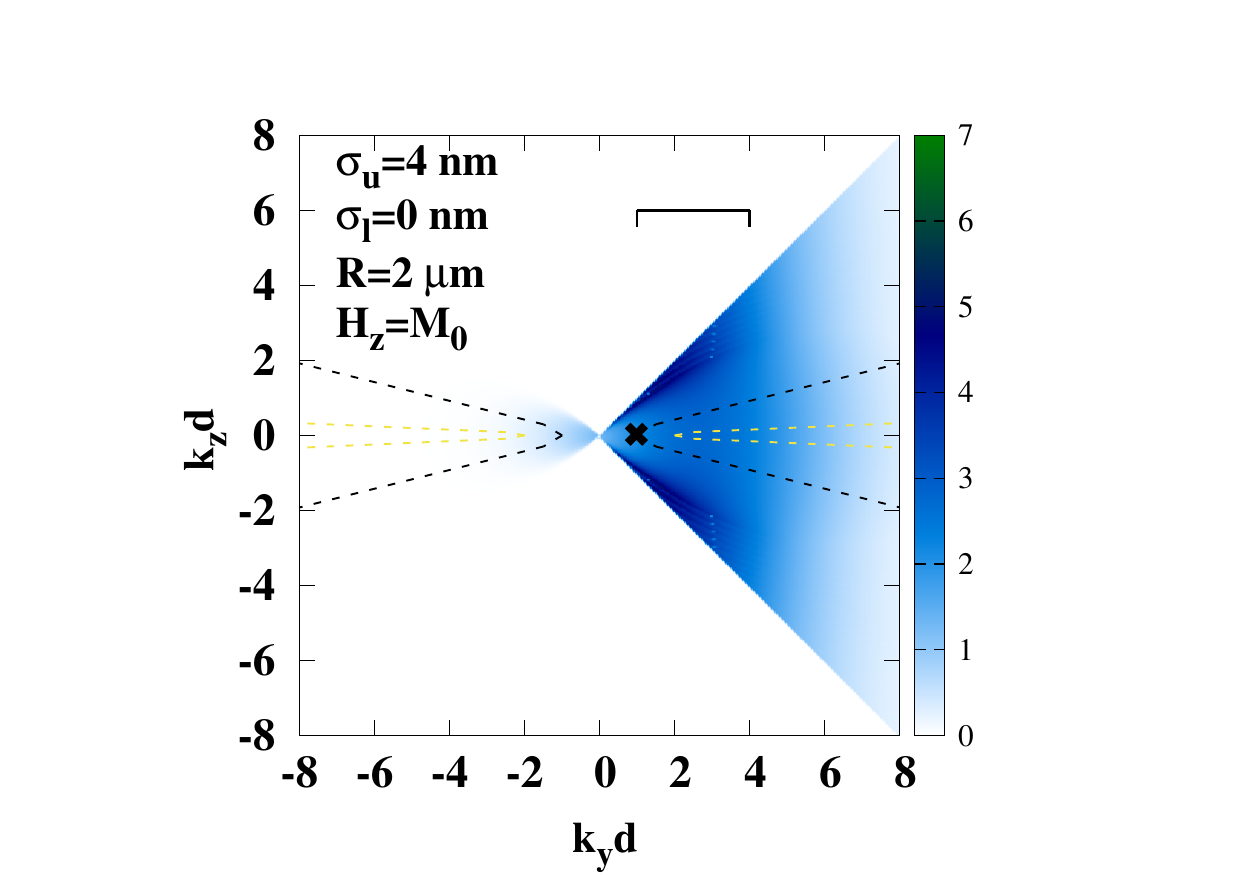}} 
\end{center}
\caption{(Color online) Momentum $\mathbf{k}$ dependence of the scattering
potential $|\mathcal{B}_{\mathbf{k}\mathbf{k^{\prime}}}|$ (in units of $
10^{-8}\protect\mu_{0}\protect\gamma M_{0}$) between DE modes. $\mathbf{k}
^{\prime}\ $is fixed to $(1/d)\hat{\mathbf{y}}$, i.e. the cross in the
figure. The black and orange dashed curves represent the equal-frequency
contours for magnons with momentum $\mathbf{k}^{\prime}=(1/d)\hat{\mathbf{y}}
$ and $(2/d)\hat {\mathbf{y}}$, respectively. $d$ is the film thickness, $
\protect\sigma_{u/l}$ the rms amplitude (upper/lower surface), and $R$ the
correlation length of the surface roughness. The vertical bar indicates $2d/R
$. }
\label{DD_map}
\end{figure}

Fig.~\ref{DD_map} shows a plot of the effective scattering potential $| \mathcal{B}_{j\mathbf{k} j\mathbf{k^{\prime}}}|$ [defined below Eq. (\ref{TotIntHam})], where index $j$ is that of the DE band, between a DE mode with momentum $\mathbf{k}^{\prime} = (1/d)\hat{\mathbf{y}}$ and DE modes with momentum $\mathbf{k}$ for in-plane magnetic field $H_z = M_0$ and correlation function Eq.~(\ref{correlation}).  

The rough upper surface scatters magnons with positive momentum into magnons on the same surface, while backscattering to magnons on the remote surface is suppressed, for larger $k$ almost completely. The phase space for scattering is defined by the white and blue boundary $k_z = \sqrt{H_z/M_0}k_y,$ defining the degeneracy of the DE and bulk modes.\textit{\ }We observe that the scattering is dominated by small momentum transfer $ \left\vert \mathbf{k} - \mathbf{k}^{\prime}\right\vert \lesssim2/R$. Since for two-magnon scattering the frequency is conserved, we plot the iso-frequency contours for the magnons with momentum $\mathbf{k}^{\prime} = (1/d)\hat{\mathbf{y}}$ (black) and $(2/d)\hat{\mathbf{y}}$ (orange) respectively, illustrating that with larger momentum the magnons are increasingly scattered in the forward direction, reflecting the \textquotedblleft ridge\textquotedblright-like energy spectra of DE magnons \cite{DE}. This feature allows simplifications of the analysis of DE magnon surface damping and transport (see Sec.~\ref{transport}) below.  

As discussed above, DE magnons with momentum $\mathbf{k} = |k_y|\hat{\mathbf{y}}$ can only scatter into other DE magnons. We find the surface damping coefficient from the self-energy by self-consistently solving the integral equations \cite{Ando} (omitting the constant band index)  
\begin{align}
	 \Sigma_{\mathbf{k}}(\omega_{\mathbf{k}}) &= \sum_{\mathbf{k}^{\prime}}| \mathcal{B}_{\mathbf{k}\mathbf{k}^{\prime}}|^2\frac{G_{\mathbf{k}^{\prime}}^{(0)}(\omega_{\mathbf{k}})}{1 - G_{\mathbf{k}^{\prime}}^{(0)}(\omega_{\mathbf{k}})\Sigma_{\mathbf{k}^{\prime}}(\omega_{\mathbf{k}})} \notag \\
	 & \approx \sum_{\mathbf{k}^{\prime}}|\mathcal{B}_{\mathbf{k}\mathbf{k}^{\prime}}|^2\frac{G_{\mathbf{k}^{\prime}}^{(0)}(\omega_{\mathbf{k}})}{1 - G_{\mathbf{k}^{\prime}}^{(0)}(\omega_{\mathbf{k}})\Sigma_{\mathbf{k}}(\omega_{\mathbf{k}})} . \label{integral} 
\end{align} 
In the last step, we invoke the long-range nature of the scattering potential $|\mathcal{B}_{\mathbf{k}\mathbf{k}^{\prime}}|^2\propto e^{- \left\vert \mathbf{k} - \mathbf{k}^{\prime}\right\vert^2R^2/4}$ that allows us to replace the self-energy $\Sigma_{\mathbf{k}^{\prime}}(\omega_{\mathbf{k}})$ by $\Sigma_{\mathbf{k}}(\omega_{\mathbf{k}})$. Eq.~(\ref {integral}) is numerically solved by carrying out the integral of $\mathbf{k}^{\prime}$ explicitly.  

The long-range nature of the scattering potential implies localization of the scattering in momentum space with the analytical estimate  
\begin{equation}
	 \Sigma_{\mathbf{k}}(\omega_{\mathbf{k}}) \approx \frac{|\mathcal{B}_{\mathbf{k} \mathbf{k}}|^2}{i\Gamma_{\mathbf{k}} - \Sigma_{\mathbf{k}}(\omega_{\mathbf{k}})}\frac{L^2}{4\pi^2}S, 
\end{equation} 
where $L^2$ is the sample area and  
\begin{equation}
	 S \approx \frac{2\sqrt{H_z/M_0}}{2\pi}\pi\left(\frac{2}{R}\right)^2 = \frac{4\sqrt{H_z/M_0}}{R^2} 
\end{equation} 
denotes the scattering area in reciprocal space (see Fig.~\ref{DD_map}). Disregarding the small intrinsic Gilbert damping $\alpha_0$ and the real part of the self-energy, we find  
\begin{equation}
	 \left\vert \mathrm{Im}\Sigma_{\mathbf{k}}\right\vert \approx \frac{L}{\pi R}|\mathcal{B}_{\mathbf{k}\mathbf{k}}|\left(\frac{H_z}{M_0}\right)^{1/4} . \label{no_integral} 
\end{equation} 
$\left\vert \mathcal{B}_{\mathbf{k}\mathbf{k}}\right\vert \propto k\sigma R/L$  implies that ${\rm Im} \Sigma_{\mathbf{k}}\propto\sigma k_y$ but does not depend on $R$. When $H_z = M_0$, and $k_yd = 2$ ($k_yd = 3$), $ \alpha_s = 5 . 67\times10^{- 3}$ ($0 . 84\times10^{- 2}$) which is not far off the numerical results $\alpha_s = 7 . 0\times10^{- 3}$ ($1 . 24\times10^{- 2}$).  

\begin{figure}[th]
\begin{center}
{\includegraphics[width=8.5cm]{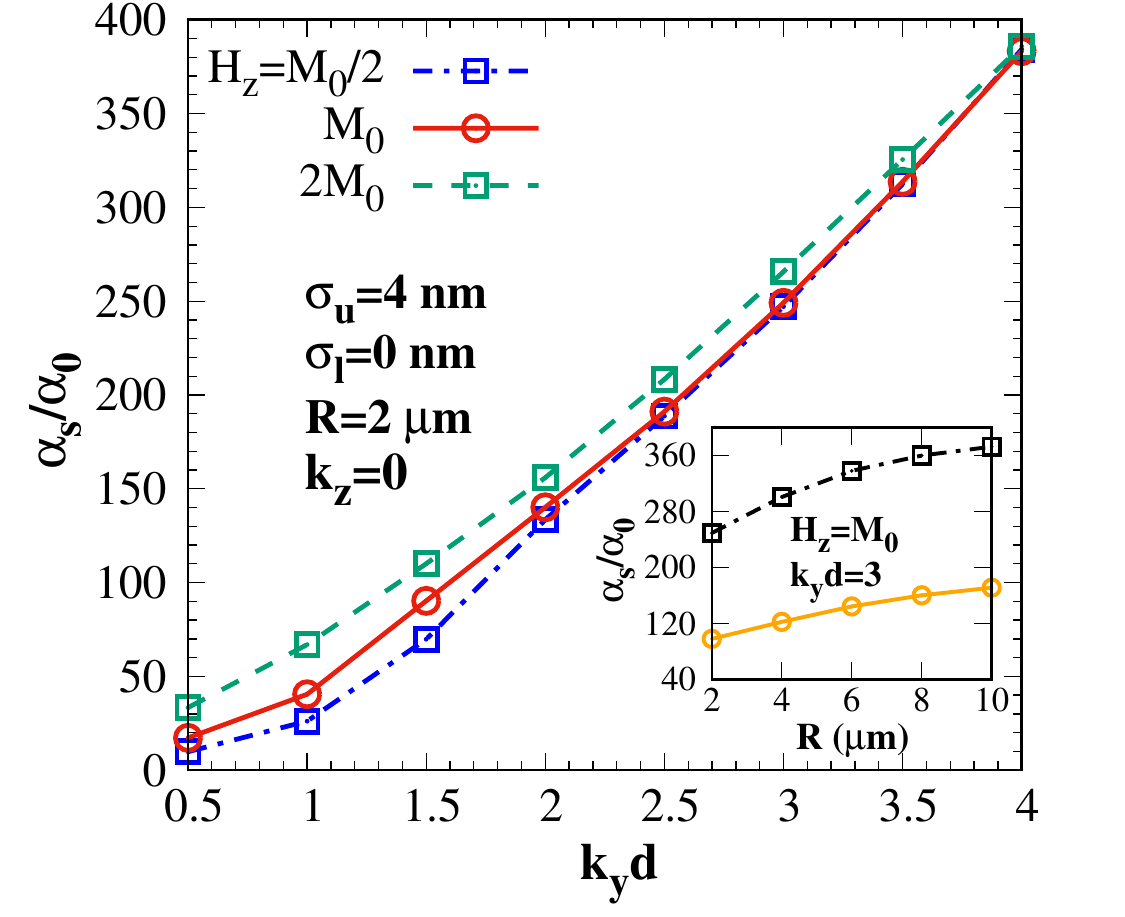}}
\end{center}
\caption{(Color online) Momentum dependence of surface damping coefficient $
\protect\alpha_{s}$ relative to the intrinsic Gilbert damping $\protect\alpha
_{0}=5\times10^{-5}$. The applied magnetic fields are $H_{z}=M_{0}/2$ (blue
dashed-dotted curve with squares), $M_{0}$ (red solid curve with circles)
and $2M_{0}$ (green dashed curve with squares), respectively. Inset:
Correlation length $R$ dependence of $\protect\alpha_{s}$ for $H_{z}=M_{0}$
and $k_{y}d=3$. The black dot-dashed curve with squares and solid curve with
circles are calculated with $\protect\sigma_{u}=4~\mathrm{\protect\mu}$
\textrm{m} and $2~\mathrm{\protect\mu}$m, respectively. }
\label{fig:surface_damping}
\end{figure}

Fig.~\ref{fig:surface_damping} is a plot of the $k_y$- and in-plane magnetic field dependence of the calculated surface damping coefficient $ \alpha_s$ that is normalized by the intrinsic Gilbert damping $ \alpha_0 = 5\times 10^{- 5}$ confirming the approximate linear dependence $ \alpha_s\propto\sigma k_y$ derived above, for larger momenta $ k_yd\gtrsim2$ and $\mathbf{k} = k_y\hat{\mathbf{y}}$. Physically, this effect is caused by the increasing localization of the wave functions to the surface which becomes more susceptible to the roughness, while simultaneously the phase space for scattering increases. The enhanced surface damping coefficient for larger wave numbers $k_yd\gtrsim2 . 5$ is of the order of $0 . 01$, much larger than the Gilbert damping in YIG, which should hinder the spectroscopic observation of DE modes \cite{Brillouin_light_scattering,Nobel,Hashimoto} as well as the manipulation of magnons by light \cite{optical_cooling}.  

At large momenta, the coupling strength between DE modes, determined by the amplitude overlap at the sample surface, $|\mathcal{B}_{\mathbf{k}\mathbf{k}^{\prime}}|^2\propto kk^{\prime}$ increases significantly [see Eqs.~(\ref{quadraticB}) and (\ref{Zeeman_quadratic})], reflecting their increased surface localization. At large momenta (or strong surface roughness), the self-consistent Born approximation breaks down \cite{Mahan}. The more involved single-site approximation could then be used \cite{Velicky}, but we note that the divergence for large wave numbers is an artifact of the magnetostatic approximation: the exchange interaction eventually adds a finite mass term \cite{Kalinikos,exchange_1969,exchange_six_order} that reduces the amplitude of DE mode at the sample surface and hence the scattering potential. A cut-off momentum $k_c$ can take care of the exchange effect as follows \cite{Kalinikos,exchange_1969,exchange_six_order} . When the exchange energy $\mu_0\gamma M_0\alpha_{\mathrm{ex}}k^2$ is one order of magnitude smaller than the dipolar one $\mu_0\gamma M_0$, i.e., $\alpha_{\mathrm{ex}}k_c^2\gtrsim0 . 1$, the exchange interaction can be disregarded. For YIG with $\alpha_{\mathrm{ex}} = 3\times10^{- 16}$ \cite{exchange_stiffness,exchange_1969,exchange_six_order}, $k_c\gtrsim5 \times10^6~\mathrm{m}^{- 1}$. With our film thickness $d = 3\times 10^{- 6}$ ~m, $k_cd\gtrsim15$. Here, we focus on momenta $kd\simeq4$, which implies still relatively weak coupling as well as absence of exchange effects.  

For DE magnons with $\mathbf{k} = k_y\hat{\mathbf{y}}$, when $k_yd\gtrsim2$ , $\alpha_s(\omega_{\mathbf{k}})\sim\sigma k_y$, and increases slowly with large $R$ when $R\gtrsim d$. The inset in Fig.~\ref{fig:surface_damping} shows these dependencies for typical parameters $H_z = M_0$ and $k_yd = 3$ . The effect of the enhanced scattering potential by a large $R$ [see Eq.~(\ref{correlation})] is largely cancelled by the simultaneous squeezing of the magnon scattering phase space (see Fig.~\ref{DD_map}). The small effect of an applied field $H_z$ is caused by another cancellation of two effects: On one hand, the effective scattering potential contributed by the Zeeman perturbation [Eq.~(\ref{Zeeman_quadratic})] is proportional to $H_z, $ while on the other hand the Lorentzian magnon spectral function broadens with $H_z$ for constant $\alpha_0$. As long as $|\mathbf{k}|d\gtrsim1,$ $ \alpha_s$ does not depend strongly on the thickness of the film either, because the surface magnon wave function $m_{x,y}^{\mathbf{k}}(d/2)\propto \sqrt{k}$ [see Eq.~(\ref{normalization2})] and hence the local scattering potentials in Eqs.~(\ref{quadraticB}) and (\ref{Zeeman_quadratic}) do not depend significantly on the thickness of the sample. Hence the surface-induced damping of surface magnons in magnetic spheres is not expected to depend on a radius in the sub-millimeter range \cite{WGM1,WGM2,WGM3,Sanchar_PRB}. Also, surface damping only weakly depends on a bulk Gilbert damping when $\alpha_0\ll\alpha_s$.  

\section{Transport of DE magnons} \label{transport}
 Forward scattering is not as harmful for transport as back scattering. Large differences in the single-particle and transport lifetimes of electrons therefore exist when the scattering potential is long-range  \cite{Abrikosov,Fetter,Mahan,Haug}. We may expect similar physics for DE-magnon transport in the linear response regime \cite{Abrikosov,Fetter,Mahan,Haug}.  

\subsection{Transport relaxation time}  

The magnon current $\mathbf{J}^m$ and heat current $\mathbf{J}_Q$ respond to the chemical potential $\mu_m$ and temperature gradients as  
\begin{equation}
	 \left(
\begin{array}
{c} \mathbf{J}_m \\
	 \mathbf{J}_Q 
\end{array} \right) = \left(
\begin{array}
{cc} \mathcal{L}^{(11)} & \mathcal{L}^{(12)} \\ \mathcal{L}^{(12)} & \mathcal{L}^{(22)} 
\end{array} \right) \left(
\begin{array}
{c} \nabla\mu_m \\ \nabla T/(k_BT) 
\end{array} \right) , 
\end{equation} 
where $\mathcal{L}^{\left(ij\right)}$ are linear response functions \cite{spin_caloritronics,cornelissen}. Here, we focus on transport by density and field gradients and $\nabla T = 0$, i.e., the magnon (spin) conductivity $ \mathcal{L}^{(11)}\equiv\mathcal{L}$. In the static limit,  
\begin{equation}
	 \mathrm{Re}\mathcal{L}_{\alpha\alpha} =  - \mathrm{\lim_{\omega\rightarrow0}{\rm Im}}\frac{\Pi_{\alpha\alpha}^{\mathrm{ret}}(\omega)}{\omega}, 
\end{equation} 
with $\alpha = \{y,z\}$.  
\begin{equation}
	 \Pi_{\alpha\alpha}^{\mathrm{ret}}(\omega) =  - i\int_{- \infty}^{\infty}dt^{\prime}\Theta(t - t^{\prime})e^{i\omega(t - t^{\prime})}\left\langle [\hat{\jmath}_{\alpha}^{\dagger}(t),\hat{\jmath}_{\alpha}(t^{\prime})]\right \rangle \label{correlation2} 
\end{equation} 
is the retarded current-current correlation function. $\hat{\jmath}_{\alpha} = \sum_{\mathbf{k}}v_{\mathbf{k}}^{\alpha} \hat{\alpha}_{\mathbf{k}}^{\dagger}\hat{\alpha}_{\mathbf{k}}$ represents the magnon current operator in terms of the magnon group velocity $\mathbf{v}_{\mathbf{k}}\equiv\partial\omega_{\mathbf{k}}/\partial\mathbf{k}$. For DE magnons with momentum $\mathbf{k} = k_y\hat{\mathbf{y}}$,  
\begin{equation}
	 v_{k_y}^y = \frac{(\mu_0\gamma M_0)^2d}{4\omega_{k_y}}e^{- 2k_yd} \label{velocity} 
\end{equation} 
with the frequency \cite{DE,new_book} \[ \omega_{k_y} = \sqrt{\omega_H^2 + \omega_H \omega_M + \omega_M^2 \frac{1 - e^{- 2k_yd}}{4} }.\] $v_{k_y}^y$ exponentially tends to zero with increasing $k_y$.  

It is again convenient to calculate first the Matsubara Green function $ \Pi_{\alpha\alpha}(i\omega_n)$ followed by analytical continuation $ i\omega_n\rightarrow\omega + i\delta$ \cite{Abrikosov,Fetter,Mahan,Haug}. Then  
\begin{equation}
	 \mathrm{Re}\ \mathcal{L}_{\alpha\alpha} = \int_{- \infty}^{\infty}\frac{d\varepsilon}{2\pi}\left(- \frac{dn_B(\varepsilon)}{d\varepsilon}\right) P(\varepsilon - i\delta,\varepsilon + i\delta), 
\end{equation} 
where $n_B(\varepsilon)\equiv(e^{\beta\varepsilon} - 1)^{- 1}$ and  
\begin{equation}
	 P(\varepsilon - i\delta,\varepsilon + i\delta) = \sum_{\mathbf{k}}v_{\mathbf{k}}^{\alpha}\Gamma_{\mathbf{k}}^{\alpha}(\varepsilon - i\delta,\varepsilon + i\delta)G_{\mathbf{k}}(\varepsilon + i\delta)G_{\mathbf{k}}(\varepsilon - i\delta) . \label{PP} 
\end{equation} 
Here, $\Gamma_{\mathbf{k}}^{\alpha}$ is the vertex function, which in the ladder approximation satisfies the integral equation  
\begin{align}
	 \Gamma_{\mathbf{k}}^{\alpha}(\varepsilon - i\delta,\varepsilon + i\delta) &= v_{\mathbf{k}}^{\alpha} + \sum_{\mathbf{k}^{\prime}}\Gamma_{\mathbf{k}^{\prime}}^{\alpha}(\varepsilon - i\delta,\varepsilon + i\delta)|\mathcal{B}_{\mathbf{k} \mathbf{k}^{\prime}}|^2 \notag \\
	 & \times G_{\mathbf{k}^{\prime}}(\varepsilon + i\delta)G_{\mathbf{k}^{\prime}}(\varepsilon - i\delta) . \label{vertex} 
\end{align} 
This integral equation is difficult to solve in general \cite{Abrikosov,Fetter,Mahan,Haug}. However, for DE magnons with momentum perpendicular to the magnetization, we can find an approximate solution for their transport perpendicular to the magnetization, i.e., $L_{yy}$, with long-ranged surface roughness as follows.  

We use the identity $G_{\mathbf{k}}(\varepsilon + i\delta)G_{\mathbf{k}}(\varepsilon - i\delta) ={A_{\mathbf{k}}(\varepsilon)}/[{2\Delta_{\mathbf{k}}(\varepsilon)}]$ with spectral function  
\begin{equation}
	 A_{\mathbf{k}}(\varepsilon) = \frac{2\Delta_{\mathbf{k}}(\varepsilon)}{(\varepsilon - \omega_{\mathbf{k}} -{\rm Re}\Sigma_{\mathbf{k}}(\varepsilon))^2 + \Delta_{\mathbf{k}}^2(\varepsilon)} 
\end{equation} 
and $\Delta_{\mathbf{k}}(\varepsilon) =  - \mathrm{Im}\Sigma_{\mathbf{k}}(\varepsilon)$ being the total broadening by the intrinsic Gilbert damping and surface roughness [see Eq.~(\ref{imaginary_self})]. The spectral function appears in both Eqs.~(\ref{PP}) and (\ref{vertex}), indicating that  $\omega_{\mathbf{k}} \approx \omega_{\mathbf{k}^{\prime}} \approx \varepsilon$ when the broadening is small and $A_{\mathbf{k}}(\varepsilon)\rightarrow2\pi \delta(\varepsilon - \omega_{\mathbf{k}})$. Both $\mathbf{k}$ and $\mathbf{k}^{\prime}$ are close to normal to the magnetization, as established in the previous sections (see Fig.~\ref{DD_map}). In other words, the DE magnons with momenta $\mathbf{k}$ are scattered mainly along the $\hat{\mathbf{y}}$-direction. Furthermore, for smooth surface roughness, the momentum transfer between DE modes is not very large. When writing $\Gamma_{\mathbf{k}}^y(\varepsilon - i\delta ,\varepsilon + i\delta) = v_{\mathbf{k}}^y\gamma_{\mathbf{k}}(\varepsilon - i\delta,\varepsilon + i\delta)$, for $| \mathbf{k}|d\gtrsim1$, using $v_{\mathbf{k}}^y\sim e^{- 2k_yd}$ from Eq.~(\ref{velocity}) and expressing $|\mathcal{B}_{\mathbf{k}\mathbf{k}^{\prime}}|^2\sim Qk_yk_y^{\prime}e^{- |\mathbf{k}^{\prime} - k_y\hat{\mathbf{y}}|^2R^2/4}$ for nearly one-dimensional scattering,  
\begin{align}
	 F &= \lim_{\varepsilon\rightarrow\omega_{\mathbf{k}}}\sum_{\mathbf{k}^{\prime}}\frac{v_{\mathbf{k}^{\prime}}^y}{v_{k_y}^y}\frac{|\mathcal{B}_{\mathbf{k}\mathbf{k}^{\prime}}|^2}{(\omega - \omega_{\mathbf{k}^{\prime}})^2 + (\Delta_{\mathbf{k}^{\prime}}(\varepsilon))^2} \notag \\
	 & \rightarrow\sum_{\mathbf{k}^{\prime}}e^{- 2(k_y^{\prime} - k_y)d}e^{- | \mathbf{k}^{\prime} - k_y\hat{\mathbf{y}}|^2R^2/4}\frac{Qk_yk_y^{\prime}}{(\omega_{\mathbf{k}} - \omega_{\mathbf{k}^{\prime}})^2 + \Delta_{\mathbf{k}}^2} . \label{suppression} 
\end{align} 
The first and second exponentials limit the scattering vectors $\left\vert k_y^{\prime} - k_y\right\vert \lesssim1/(2d)$ and $\left\vert k_y^{\prime} - k_y\right\vert \lesssim2/R$, respectively. When $ 2/R\lesssim1/(2d)$ and hence $R\gtrsim4d$, substituting the \textquotedblleft mean value\textquotedblright\ of $k_y^{\prime}$ by $ k_y + 1/R$ in the first exponential leads to  
\begin{equation}
	 F\gtrsim e^{- 2d/R}\sum_{\mathbf{k}^{\prime}}\frac{|\mathcal{B}_{\mathbf{k} \mathbf{k}^{\prime}}|^2}{(\omega_{\mathbf{k}} - \omega_{\mathbf{k}^{\prime}})^2 + \Delta_{\mathbf{k}}^2} = e^{- 2d/R}, \label{suppression2} 
\end{equation} 
where we used  
\begin{equation}
	 \sum_{\mathbf{k}^{\prime}}|\mathcal{B}_{\mathbf{k}\mathbf{k}^{\prime}}|^2 \frac{1}{(\omega_{\mathbf{k}} - \omega_{\mathbf{k}^{\prime}})^2 + ({\rm Im} \Sigma_{\mathbf{k}})^2} = 1 \label{hard} 
\end{equation} 
from the self-consistent Born approximation. ${\gamma}_{\mathbf{k}}$ therefore does not depend on $\mathbf{k}$ to leading order when $R\gtrsim4d$ . This allows application of the mean value theorem which leads to $\gamma_{\mathbf{k}^{\prime}} \approx \gamma_{\mathbf{k}}$. We arrive at the closed expression  
\begin{equation}
	 \Gamma_{\mathbf{k}}^y(\varepsilon - i\delta,\varepsilon + i\delta) \approx v_{\mathbf{k}}^y\left(1 - F\right)^{- 1}, 
\end{equation} 
and  
\begin{equation}
	{\rm Re}\mathcal{L}_{yy} = \int_{- \infty}^{\infty}\frac{d\varepsilon}{2\pi} \left(- \frac{dn_B(\varepsilon)}{d\varepsilon}\right) \sum_{\mathbf{k}}(v_{\mathbf{k}}^y)^2\frac{A_{\mathbf{k}}(\varepsilon)}{2\Delta_{\mathbf{k}}^t(\varepsilon)}, \label{conductivity} 
\end{equation} 
where  
\begin{equation}
	 2\Delta_{\mathbf{k}}^t(\varepsilon) = 2\Delta_{\mathbf{k}}(\varepsilon)\left[ 1 - \sum_{\mathbf{k}^{\prime}}\frac{v_{\mathbf{k}^{\prime}}^y}{v_{\mathbf{k}}^y}\frac{A_{\mathbf{k}^{\prime}}(\varepsilon)}{2\Delta_{\mathbf{k}^{\prime}}(\varepsilon)}|\mathcal{B}_{\mathbf{k}\mathbf{k}^{\prime}}|^2 \right] . \label{broadening_tr} 
\end{equation} 
We thus derived a relation between the lifetime broadening in Eq.~(\ref {broadening_tr}) and the transport damping coefficient for the magnon propagating nearly perpendicular to the magnetization:  
\begin{equation}
	 \alpha_t(\omega_{\mathbf{k}}) = \frac{2\Delta_{\mathbf{k}}^t}{\omega_{\mathbf{k}}} = \alpha_s(\omega_{\mathbf{k}})\left(1 - F\right) . \label{transport_damping} 
\end{equation}

With $d = 3~\mathrm{\mu m}$, the suppressing factor Eq.~(\ref{suppression}) is calculated to be $F\gtrsim0 . 61$ when $R = 12$~$\mathrm{\mu m}$, and $ F\gtrsim0 . 74$ when $R = 20$~$\mathrm{\mu m}$. $\alpha_s$ does not change much with larger $R$ when $k_yd\gtrsim2,$ see Fig.~\ref {fig:surface_damping}, and hence $\alpha_t$ decreases exponentially with increasing $R$. The transport of DE magnons perpendicular to the magnetization is therefore efficient for smooth surface roughness, i.e., when $R\gtrsim4d$, even though their lifetime can be very short. For larger $ \mathbf{k}$ or shorter-ranged roughness, i.e. $R\lesssim4d$, $ \alpha_t(\omega_{\mathbf{k}})\lesssim \alpha_s(\omega_{\mathbf{k}})$ still holds, but the transport of DE magnons is not protected anymore because the group velocity and in-scattering of DE magnons exponentially decreases. We conclude that smooth surface roughness affects the transport of DE magnons much less than the large lifetime broadening suggests, which is caused by chirality and long-range disorder, which both favor strong forward scattering.  

\subsection{Chiral conductivity}  

As addressed in Sec.~\ref{numerical_damping}, DE magnons propagating in opposite directions experience different scattering potential when the surface roughness is different at the two surfaces, which leads to different magnon conductivities when the in-plane magnetic field is reversed, i.e., $ L_{yy}^{ij}(\mathbf{M})\neq L_{yy}^{ij}(- \mathbf{M})$. The spin conductivity can be calculated or estimated from Eq.~(\ref{conductivity}). In the weak scattering regime, the spectral function $A_{\mathbf{k}}(\varepsilon)\rightarrow2\pi\delta(\varepsilon - \omega_{\mathbf{k}})$, and the spin conductivity reduces to the conventional form from the Boltzmann equation  \cite{cornelissen,Boltzmann},  
\begin{equation}
	 \mathcal{L}_{yy} ={\rm Re}\mathcal{L}_{yy}^{\left(11\right)} = \sum_{\mathbf{k}}(v_{\mathbf{k}}^y)^2\frac{1}{2\Delta_{\mathbf{k}}^t}\left(- \frac{dn_B(\omega_{\mathbf{k}})}{d\omega_{\mathbf{k}}}\right) , \label{spin_conductivity} 
\end{equation} 
where $n_B$ is the Boltzmann distribution function. The spin Seebeck coefficient $\mathcal{L}^{\left(12\right)}$ and magnon heat conductivity $ \mathcal{L}^{\left(22\right)}$ are obtained by replacing one or two magnon number-current operators $\hat{\jmath}_{\alpha}$ in Eq.~(\ref{correlation2}) by the magnon energy-current operator $\hat{\jmath}_{\alpha}^Q = \sum_{\mathbf{k}}\hbar\omega_{\mathbf{k}}v_{\mathbf{k}}^{\alpha}\hat{\alpha}_{\mathbf{k}}^{\dagger}\hat{\alpha}_{\mathbf{k}}$ \cite{Mahan}, leading to  \cite{Mahan,cornelissen,Boltzmann}  
\begin{align}
	{\rm Re}L_{yy}^{12} &= \sum_{\mathbf{k}}(v_{\mathbf{k}}^y)^2\frac{\hbar\omega_{\mathbf{k}}}{2\Delta_{\mathbf{k}}^t}\left(- \frac{dn_B(\omega_{\mathbf{k}})}{d\omega_{\mathbf{k}}}\right) \approx \hbar\omega_{\mathrm{DE}}L_{yy}, \label{spin Seebeck} \\
	{\rm Re}L_{yy}^{22} &= \sum_{\mathbf{k}}(v_{\mathbf{k}}^y)^2\frac{(\hbar\omega_{\mathbf{k}})^2}{2\Delta_{\mathbf{k}}^t}\left(- \frac{dn_B(\omega_{\mathbf{k}})}{d\omega_{\mathbf{k}}}\right) \approx (\hbar\omega_{\mathrm{DE}})^2L_{yy} . \label{heat_conductivity} 
\end{align} 
where the approximation is allowed when limiting attention to the DE magnons with narrow band width \cite{DE}. $v_{\mathbf{k}}^y$ can be estimated by Eq.~(\ref{velocity}) due to the \textquotedblleft ridge"-like shape of the DE dispersion \cite{DE}.  

\begin{figure}[th]
\begin{center}
{\includegraphics[width=8.8cm]{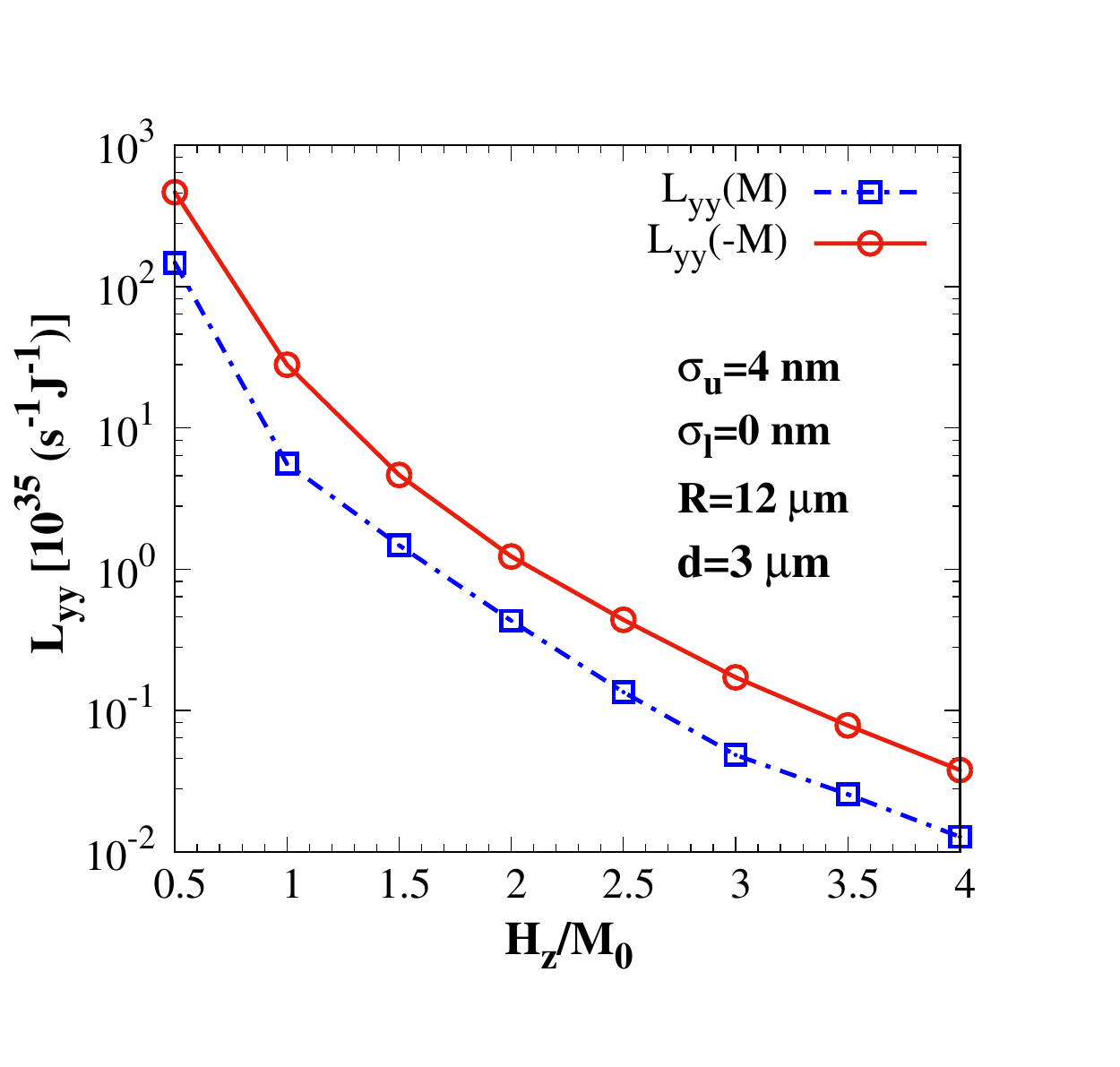}}
\end{center}
\caption{(Color online) Magnetic-field dependence of spin conductivities $
L_{yy}(\mathbf{M})$ (blue dashed curve with squares) and $L_{yy}(-\mathbf{M})
$ (red solid curve with circles) at the room temperature $T=300$~K.}
\label{chiral_conductivity}
\end{figure}

Fig.~\ref{chiral_conductivity} shows the magnetic-field dependence of the magnon conductivities $L_{yy}(\mathbf{M})$ and $L_{yy}(- \mathbf{M})$ at room temperature $T = 300$~K. When the upper surface is rough with $\sigma_u = 4~ \mathrm{\mu m}$ and $R = 12~\mathrm{\mu m}$ but the lower surface is flat, we find $L_{yy}(- \mathbf{M}) \approx 3L_{yy}(\mathbf{M})$ in a YIG film with thickness $d = 3~\mathrm{\mu m}$, where $L_{yy}(\mathbf{M})$ and $L_{yy}(- \mathbf{M})$ are dominated by the DE magnons near the upper and lower surfaces, respectively. For momenta $\left\vert \mathbf{k}\right\vert d\gtrsim 1$, the scattering is chiral, so the upper surface roughness efficiently scatters the DE magnons near the upper surface, but does not affect the modes on the lower surface. Therefore, the spin conductivity changes when the in-plane magnetic field is reversed. However, we do not generate a short circuit even in the absence of all scattering at the lower surface since the DE magnons with relatively small momenta $\left\vert \mathbf{k}\right\vert d\lesssim 1$ on the lower surface are still scattered by the upper surface roughness. $L_{yy}$ decreases with increasing magnetic field because $\omega_{\mathbf{k}}$ increases and the freeze-out effect $ dn_B(\omega_{\mathbf{k}})/d\omega_{\mathbf{k}}\propto 1/\omega_{\mathbf{k}}^2$. From the transport lifetime $\tau_{\mathbf{k}}^t\equiv 1/(2\Delta_{\mathbf{k}}^t)$, we expect $L_{yy}^{- 1}\propto \sigma (1 - e^{- 2d/R})$.   

We may compare the surface conducticivity with that of the parallel channel of the bulk exchange modes with higher energy but larger group velocity. From the calculated bulk conductivity $L_b$ at room temperature, in the film with $d = 3~\mathrm{\mu m,}$ $L^{\prime} = L_bd \approx 5\times 10^{41}~({\rm s}\cdot{\rm J})^{- 1}$ \cite{Boltzmann}, about four orders in magnitude larger than the surface contribution. The spin conductivity contributed by the magnetostatic bulk magnons should be much smaller than $L^{\prime}$ because they their small group velocity. DE magnon channels can still be identified in transport by their chirality or by selective excitation.  

Pirro \textit{et al}. \cite{backscattering_immune} report numerical simulations for a single strongly scattering local defect in ultrathin films and a suppression of back scattering of magnons in the DE configuration far into the exchange regime. The conclusion is similar to ours, and the physical origin may be the same, but it is difficult to compare these two very different approaches. First, this study does not address the magnon lifetime or self-energy, which is important for experiments that study their spectral properties. Next, we are able to treat thick films in which the surface states are well developed, which are difficult to model by micromagnetism. We also focus on weak long-range correlated disorder in order to exclude scattering into volume exchange modes, which reduce transport significantly when the spectra of surface and bulk modes overlap  \cite{backscattering_immune}. We plan to extend the present quasi-analytical method to assess the thin-film regime and short range scattering potentials by including the exchange interaction in a future study.  

\section{Excitation of surface magnons from surface roughness} \label{excitation}
For long-range disorder the scattering of DE magnons with momenta $k_y\hat{\mathbf{y}}$ into bulk states close to the Kittel mode is very inefficient, and we disregarded it completely in the discussion of the DE magnon lifetimes. Also for the surface conductivity, the scattering into the Kittel mode contributes only in a very small region of momentum space. However, the inverse process, i.e., the scattering of bulk magnons into surface modes with finite $\delta k_z$ is allowed (see Fig.~\ref{map_DB})  \cite{DE}: The DE modes with momenta $\mathbf{k} = k_y\hat{\mathbf{y}}$ are well separated in energy and therefore cannot scatter elastically into bulk modes. On the other hand, the DE modes very close to the boundary between bulk and DE modes with significant $\delta k_z$ are nearly degenerate with the Kittel mode \cite{DE} and can be populated via surface roughness when the latter is excited by a \textit{uniform} microwave field. DE magnon numbers on both sides of a film excited by the uniform microwave field differ when the roughness is asymmetric \cite{heatconveyer2,heatconveyer3,heatconveyer4}.   

\subsection{Model}  

We consider a subspace consisting of the Kittel modes and DE modes with momenta $\mathbf{k} = \left(0,k_y,\delta k_z\right) $, with operators $ \hat{\alpha}_K$ and $\hat{\alpha}_{\mathbf{k}}$, respectively, and interaction matrix elements $\mathcal{B}_{\mathbf{k}K}$. The Hamiltonian of non-interacting \cite{Suhl_nonlinear,nonlinear2} magnons coupled to a uniform linearly polarized microwave field $H_x\mathbf{\hat{x}}$ with frequency $\omega_d$ reads  
\begin{align}
	 \hat{H} &= \omega_K\hat{\alpha}_K^{\dagger}\hat{\alpha}_K + \sum_{\mathbf{k}}\omega_{\mathbf{k}}\hat{\alpha}_{\mathbf{k}}^{\dagger}\hat{\alpha}_{\mathbf{k}} + \sum_{\mathbf{k}}(\mathcal{B}_{\mathbf{k}K}\hat{\alpha}_K^{\dagger}\hat{\alpha}_{\mathbf{k}} \notag \\
	 & + \mathcal{B}_{\mathbf{k}K}^{\ast}\hat{\alpha}_{\mathbf{k}}^{\dagger}\hat{\alpha}_K) + 2g\hat{H}_x(t)\left(\hat{\alpha}_K + \hat{\alpha}_K^{\dagger}\right) . 
\end{align} 
Here, $\hat{H}_x(t) = \hat{h}_x(0)e^{- i\omega_dt} + \hat{h}_x^{\dagger}(0)e^{i\omega_dt}$ is the magnetic-field operator in terms of photon operator $\hat{h}_x$, and $g = \mu_0d\sqrt{2\gamma M_0/2}m_x^K$ arises from the Zeeman coupling between the Kittel mode and the uniform microwave magnetic field. The master equations for the magnon operators are obtained from the Heisenberg equation \cite{nonlinear2,input_output1,input_output2}, augmented by the dampings $\Gamma_K$ and $\Gamma_{\mathbf{k}}$:  
\begin{align}
	 \frac{d\hat{\alpha}_K}{dt} &=  - i\omega_K\hat{\alpha}_K - \Gamma_K\hat{\alpha}_K - i\sum_{\mathbf{k}}\mathcal{B}_{\mathbf{k}K}\hat{\alpha}_{\mathbf{k}} - ig\hat{H}_x(t), \label{coupled_equations1} \\
	 \frac{d\hat{\alpha}_{\mathbf{k}}}{dt} &=  - i\omega_{\mathbf{k}}\hat{\alpha}_{\mathbf{k}} - \Gamma_{\mathbf{k}}\hat{\alpha}_{\mathbf{k}} - i\mathcal{B}_{\mathbf{k}K}^{\ast}\hat{\alpha}_K . \label{coupled_equations2} 
\end{align} 
From Eq.~(\ref{coupled_equations2}), we obtain \cite{nonlinear2,input_output1,input_output2}  
\begin{align}
	 \hat{\alpha}_{\mathbf{k}}(t) &= \hat{\alpha}_{\mathbf{k}}(0)e^{- i\omega_{\mathbf{k}}t - \Gamma_{\mathbf{k}}t} \notag \\
	 & - i\mathcal{B}_{\mathbf{k}K}^{\ast}\int_0^td\tau e^{- (i\omega_{\mathbf{k}} + \Gamma_{\mathbf{k}})(t - \tau)}\hat{\alpha}_K(\tau) . 
\end{align} 
When the damping and excitation of the Kittel mode is weak, the evolution of  $\hat{\alpha}_K$ is free $d\hat{\alpha}_K/dt \approx  - i\omega_K\hat{\alpha}_K \approx  - i\omega_d\hat{\alpha}_K$ for the small time interval $\Gamma_{\mathbf{k}}$. Treating $\hat{\alpha}_K$ in a \textquotedblleft Markov approximation\textquotedblright\  \cite{input_output1,input_output2}    
\begin{equation}
	 \hat{\alpha}_K(\tau) \approx \hat{\alpha}_K(t)e^{i\omega_d(t - \tau)}, 
\end{equation}
inside the integral. We obtain at large times,
\begin{equation}
	 \hat{\alpha}_{\mathbf{k}}(t) = \hat{\alpha}_{\mathbf{k}}(0)e^{- i\omega_{\mathbf{k}}t - \Gamma_{\mathbf{k}}t} + i\mathcal{B}_{\mathbf{k}K}^{\ast}\hat{\alpha}_K(t)\frac{1 - e^{(i\omega_d - i\omega_{\mathbf{k}} - \Gamma_{\mathbf{k}})t}}{i\omega_d - i\omega_{\mathbf{k}} - \Gamma_{\mathbf{k}}}, 
\end{equation} 
which at large times settles into the steady state  
\begin{equation}
	 \hat{\alpha}_{\mathbf{k}}(t\rightarrow \infty) = \frac{- \mathcal{B}_{\mathbf{k}K}^{\ast}\hat{\alpha}_K(t\rightarrow \infty)}{\omega_{\mathbf{k}} - \omega_d - i\Gamma_{\mathbf{k}}} . \label{relation_2} 
\end{equation} 
By substituting this into Eq.~(\ref{coupled_equations1}) when $t\rightarrow \infty $:  
\begin{align}
	 \frac{d\hat{\alpha}_K}{dt} &=  - i\omega_K\hat{\alpha_K} - \Gamma_K \hat{\alpha}_K - ig\hat{H}_x(t) \notag \\
	 & \mbox{} + \sum_{\mathbf{k}}\frac{\left\vert \mathcal{B}_{\mathbf{k} K}\right\vert^2\hat{\alpha}_K}{- i(\omega_{\mathbf{k}} - \omega_d) - \Gamma_{\mathbf{k}}} . \label{Kittel_effective} 
\end{align} 
Using the rotating wave approximation \cite{nonlinear2,input_output1,input_output2},  
\begin{equation}
	 \hat{\alpha}_K(t\rightarrow \infty) = \frac{- g\hat{h}_x(0)e^{- i\omega_dt}}{\omega_K - \omega_d - i\Gamma_K - \sum_{\mathbf{k}}\frac{\left\vert \mathcal{B}_{\mathbf{k}K}\right\vert^2}{\omega_{\mathbf{k}} - \omega_d - i\Gamma_{\mathbf{k}}}} . 
\end{equation} 
From Eq.~(\ref{relation_2}), the excited DE magnon population  
\begin{equation}
	 \delta n_{\mathrm{DE}}\equiv \sum_{\mathbf{k}}\langle{\hat{\alpha}_{\mathbf{k}}^{\dagger}\hat{\alpha}_{\mathbf{k}}}\rangle = \rho_s\langle \hat{\alpha}_K^{\dagger}\hat{\alpha}_K\rangle , \label{efficiency} 
\end{equation} 
where 
\begin{equation}
	 \rho_s\equiv \sum_{\mathbf{k}}\frac{|\mathcal{B}_{\mathbf{k}K}|^2}{(\omega_{\mathbf{k}} - \omega_K)^2 + \Gamma_{\mathbf{k}}^2} \label{rhos} 
\end{equation} 
is the FMR excitation efficiency of the DE magnons.  

\subsection{Results}  

\begin{figure}[th]
\begin{center}
{\includegraphics[width=10.5cm]{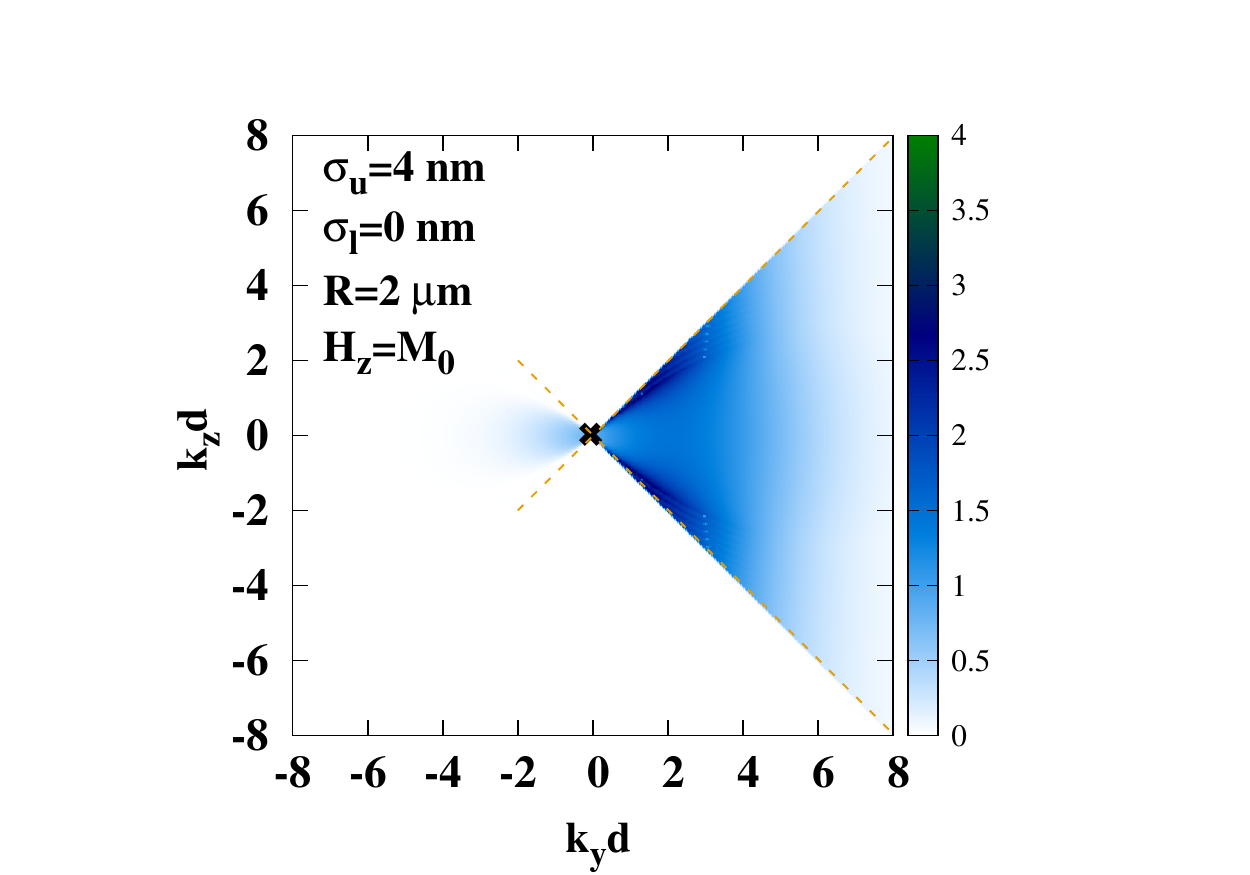}}
\end{center}
\caption{(Color online) Momentum dependence of the scattering potential $|
\mathcal{B}_{\mathbf{k}K}|$ (in units of $10^{-8}\protect\mu_{0}\protect
\gamma M_{0}$) between the Kittel mode (marked by a cross) and DE modes with
wave vector $\mathbf{k}$. The orange dashed curves $k_{z}=\pm\protect\sqrt{
H_{z}/M_{0}}k_{y}$ is the equal-frequency contour of the DE and Kittel modes
that define the boundary between surface and bulk modes \protect\cite{DE}. }
\label{map_DB}
\end{figure}

We computed the surface-roughness--assisted excitation of the DE magnons for YIG films with material parameters introduced in Sec.~\ref{numerical_damping} . The disorder on the upper and lower surfaces are chosen to be asymmetric $ \sigma_u = 4$~nm and $\sigma_l = 0$~nm and correlation length $R = 2$\thinspace $\mathrm{\mu}$m, as above. In Fig.~\ref{map_DB}, we plot the effective scattering potential $|\mathcal{B}_{\mathbf{k}K}|\ $between the Kittel mode and DE modes with momentum $\mathbf{k}$ for $H_z = M_0$.  

The Kittel mode couples dominantly with the DE modes with positive $k_y$, i.e., the ones propagating on the upper surface that is chosen to be rough, even though the microwave field is uniform \cite{heatconveyer2,heatconveyer3,heatconveyer4}. The orange dashed lines $ k_z = \pm\sqrt{H_z/M_0}k_y$ are the equal-frequency contours of the DE and Kittel modes that separate bulk and surface modes \cite{DE}.  

The efficiency $\rho_s$ of the surface-roughness--assisted excitation of DE magnons in Eq.~(\ref{rhos}) with the resonant excitation of Kittel mode $ \omega_d = \omega_K$ is $\rho_s = 2 . 4\%$, $3 . 4\%$, $4 . 8\%$ and $7 . 5\%$ for  $H_z = 0 . 5M_0$, $M_0$, 1.5$M_0$, and 2$M_0$, respectively. A significant number of DE magnons is excited during FMR and it increases with magnetic field. The excitation efficiency can be enhanced by rougher surfaces.  

The FMR-excited DE magnon with momentum $|\mathbf{k}|d\lesssim1$ are distributed by $|\mathcal{B}_{\mathbf{k}K}|^2/\left[ (\omega_{\mathbf{k}} - \omega_K)^2 + \Gamma_{\mathbf{k}}^2\right] $. The denominator is small for the magnons close to the dark-blue regions in Fig.~\ref{map_DB}. These magnons are well localized to the film surface even when $|\mathbf{k} |d\lesssim1$ with finite $\delta k_z$ and these modes are still chiral  \cite{DE}, which implies that with asymmetric surface roughness one surface is preferentially excited. These results can be tested by Brillouin light scattering spectra for films with different roughness and help to understand the heat conveyer effect \cite{heatconveyer1} in recent experiments in which a uniform magnetic field was shown to generate chiral heat transport \cite{heatconveyer2,heatconveyer3,heatconveyer4}.

\section{Summary}
 \label{summary}
 
 In conclusion, we investigated the effects of long-range, static surface roughness on the damping and excitation of surface magnons in thick magnetic films with in-plane magnetic fields. We reveal an additional damping channel for the surface magnons that strongly reduces the lifetime of surface magnons with wave number $k\gtrsim d^{- 1},$ where $d$ is the film thickness, possibly far above the bulk Gilbert damping. This indicates that the spectral features of surface magnons are smeared out by surface disorder. It is also bad news for cavity optomagnonics \cite{first_2010,second,third,fourth}  with DE modes, since the strong dephasing by surface roughness suppresses the coupling to optical whispering gallery modes. On the other hand, transport of DE magnons is protected since scattering is dominantly in the forward direction, which is caused by their nearly circular polarization and uni-directional propagation. The surface roughness also mixes the Kittel and DE modes quite efficiently such that even a uniform microwave field can pump considerable amounts of surface magnons out of the magnetic order, which is observable by Brillouin light scattering experiments. Moreover, an asymmetry of the surface roughness on both sides of the film, generates unbalanced distributions of the surface magnons and chirality during spin and heat transport.  

The surface roughness may be also dynamic, i.e., is both space and time dependent generated by thermal surface acoustic waves \cite{dynamical1,dynamical2,dynamical3}. We will show in future work that our framework for the static surface roughness may be generalized to the dynamic one.  

\begin{acknowledgments}
	This work is financially supported by the Nederlandse Organisatie voor Wetenschappelijk Onderzoek (NWO) as well as JSPS KAKENHI Grant No. 26103006. One of the authors (TY) would like to thank Simon Streib for useful discussions.
\end{acknowledgments}

%\begin{appendix}

%\end{appendix}

%merlin.mbs apsrev4-1.bst 2010-07-25 4.21a (PWD, AO, DPC) hacked
%Control: key (0)
%Control: author (72) initials jnrlst
%Control: editor formatted (1) identically to author
%Control: production of article title (-1) disabled
%Control: page (0) single
%Control: year (1) truncated
%Control: production of eprint (0) enabled

%\begin{thebibliography}{99}

%review on magnonics

%\end{thebibliography}


\begin{thebibliography}{99}%
\makeatletter
\providecommand \@ifxundefined [1]{%
 \@ifx{#1\undefined}
}%
\providecommand \@ifnum [1]{%
 \ifnum #1\expandafter \@firstoftwo
 \else \expandafter \@secondoftwo
 \fi
}%
\providecommand \@ifx [1]{%
 \ifx #1\expandafter \@firstoftwo
 \else \expandafter \@secondoftwo
 \fi
}%
\providecommand \natexlab [1]{#1}%
\providecommand \enquote  [1]{``#1''}%
\providecommand \bibnamefont  [1]{#1}%
\providecommand \bibfnamefont [1]{#1}%
\providecommand \citenamefont [1]{#1}%
\providecommand \href@noop [0]{\@secondoftwo}%
\providecommand \href [0]{\begingroup \@sanitize@url \@href}%
\providecommand \@href[1]{\@@startlink{#1}\@@href}%
\providecommand \@@href[1]{\endgroup#1\@@endlink}%
\providecommand \@sanitize@url [0]{\catcode `\\12\catcode `\$12\catcode
  `\&12\catcode `\#12\catcode `\^12\catcode `\_12\catcode `\%12\relax}%
\providecommand \@@startlink[1]{}%
\providecommand \@@endlink[0]{}%
\providecommand \url  [0]{\begingroup\@sanitize@url \@url }%
\providecommand \@url [1]{\endgroup\@href {#1}{\urlprefix }}%
\providecommand \urlprefix  [0]{URL }%
\providecommand \Eprint [0]{\href }%
\providecommand \doibase [0]{http://dx.doi.org/}%
\providecommand \selectlanguage [0]{\@gobble}%
\providecommand \bibinfo  [0]{\@secondoftwo}%
\providecommand \bibfield  [0]{\@secondoftwo}%
\providecommand \translation [1]{[#1]}%
\providecommand \BibitemOpen [0]{}%
\providecommand \bibitemStop [0]{}%
\providecommand \bibitemNoStop [0]{.\EOS\space}%
\providecommand \EOS [0]{\spacefactor3000\relax}%
\providecommand \BibitemShut  [1]{\csname bibitem#1\endcsname}%
\let\auto@bib@innerbib\@empty
%</preamble>

%review on magnonics
\bibitem{magnonics1} B. Lenk, H. Ulrichs, F. Garbs, and M. M\"{u}nzenberg,
Phys. Rep. \textbf{507}, 107 (2011).

\bibitem{magnonics2} A. V. Chumak, V. I. Vasyuchka, A. A. Serga, and B.
Hillebrands, Nat. Phys. \textbf{11}, 453 (2015).

\bibitem{magnonics3} D. Grundler, Phys. Rep. \textbf{11}, 407 (2016).

\bibitem{magnonics4} V. E. Demidov, S. Urazhdin, G. de Loubens, O. Klein, V.
Cros, A. Anane, and S. O. Demokritov, Phys. Rep. \textbf{673}, 1 (2017).

%spin Seebeck effect

\bibitem{spin_caloritronics} G. E. W. Bauer, E. Saitoh and B. J. van Wees,
Nat. Mater. \textbf{11}, 391 (2012).

\bibitem{Walker} L. R. Walker, Phys. Rev. \textbf{105}, 390 (1957).

\bibitem{DE} R. W. Damon and J. R. Eshbach, J. Phys. Chem. Solids \textbf{19}%
, 308 (1961).

\bibitem{spin_waves_book} A. Akhiezer, V. Bar\'iakhtar, and S. Peletminski, 
\textit{Spin Waves} (North-Holland, Amsterdam, 1968).

\bibitem{new_book} D. D. Stancil and A. Prabhakar, \textit{Spin
	Waves--Theory and Applications} (Springer, New York, 2009). %heat conveyer

\bibitem{heatconveyer1} T. An, V. I. Vasyuchka, K. Uchida, A. V. Chumak, K.
Yamaguchi, K. Harii, J. Ohe, M. B. Jungfleisch, Y. Kajiwara, H. Adachi, B.
Hillebrands, S. Maekawa, and E. Saitoh, Nat. Mat. \textbf{12}, 549 (2013).

\bibitem{heatconveyer2} O. Wid, J. Bauer, A. M\"uller, O. Breitenstein, S.
S. P. Parkin, and G. Schmidt, Sci. Rep. \textbf{6}, 28233 (2016).

\bibitem{heatconveyer3} E. Shigematsu, Y. Ando, S. Dushenko, T. Shinjo, and
M. Shiraishi, Appl. Phys. Lett. \textbf{112}, 212401 (2018).

\bibitem{heatconveyer4} P. Wang, L. F. Zhou, S. W. Jiang, Z. Z. Luan, D. J.
Shu, H. F. Ding, and D. Wu, Phys. Rev. Lett. \textbf{120}, 047201 (2018).

%magnons_optical waves



  \bibitem{WGM1} A. Osada, R. Hisatomi, A. Noguchi, Y. Tabuchi, R. Yamazaki,
K. Usami, M. Sadgrove, R. Yalla, M. Nomura,
and Y. Nakamura, Phys. Rev. Lett. {\bf 116}, 223601 (2016).

\bibitem{WGM2} X. Zhang, N. Zhu, C.-L. Zou, and H. X. Tang, Phys. Rev.
Lett. {\bf 117}, 123605 (2016).

\bibitem{WGM3} J. A. Haigh, A. Nunnenkamp, A. J. Ramsay, and A. J.
Ferguson, Phys. Rev. Lett. {\bf 117}, 133602 (2016).

\bibitem{Sanchar_PRB} S. Sharma, Y. M. Blanter, and G. E. W. Bauer, Phys.
Rev. B \textbf{96}, 094412 (2017).

\bibitem{optical_cooling} S. Sharma, Y. M. Blanter, and G. E. W. Bauer,
Phys. Rev. Lett. \textbf{121}, 087205 (2018).

%static surface roughness

\bibitem{static1} M. Sparks, R. Loudon, and C. Kittel, Phys. Rev. \textbf{122%
}, 791 (1961).

\bibitem{static3} R. Arias and D. L. Mills, Phys. Rev. B \textbf{60}, 7395
(1999).

\bibitem{static4} A. Y. Dobin and R. H. Victora, Phys. Rev. Lett \textbf{92}%
, 257204 (2004).

\bibitem{static2} E. Schl\"omann, J. Appl. Phys. \textbf{41}, 1617 (1969).

\bibitem{backscattering_immune} M. Mohseni, T. Bracher, Q. Wang, D. A.
Bozhko, R. Verba, B. Hillebrands, and P. Pirro, arXiv:1806.01554.

\bibitem{Tao_MoS2} T. Yu and M. W. Wu, Phys. Rev. B \textbf{93}, 045414
(2016).

\bibitem{Landau} L. D. Landau and E. M. Lifshitz, \textit{Electrodynamics of
	Continuous Media}, 2nd ed. (Butterworth-Heinenann, Oxford, 1984).

%magnetizations

\bibitem{magnetization1} A. G. Gurevich and G. A. Melkov, \textit{%
	Magnetization Oscillations and Waves} (CRC, Boca Raton, FL, 1996).

\bibitem{magnetization2} P. Hansen, J. Appl. Phys. \textbf{45}, 3638 (1974).

%exchange stiffness

\bibitem{exchange_stiffness} S. Klingler, A. V Chumak, T. Mewes, B.
Khodadadi, C. Mewes, C. Dubs, O. Surzhenko, B. Hillebrands, and A. Conca, J.
Phys. D \textbf{48}, 015001 (2015).

%wavefunction, normalization

\bibitem{magnetic_nanodots} R. Verba, G. Melkov, V. Tiberkevich, and A.
Slavin, Phys. Rev. B \textbf{85}, 014427 (2012).

%Bogoliubov

\bibitem{Kittel_book} C. Kittel, \textit{Quantum Theory of Solids} (Wiley,
New York, 1963).

\bibitem{squeezed_magnon} A. Kamra and W. Belzig, Phys. Rev. Lett. \textbf{%
	116}, 146601 (2016).

\bibitem{HP} T. Holstein and H. Primakoff, Phys. Rev. \textbf{58}, 1098
(1940).

\bibitem{Kalinikos} B. A. Kalinikos, Sov. J. Phys. \textbf{24}, 718 (1981).

%nonliear magnons

\bibitem{Suhl_nonlinear} H. Suhl, J. Phys. Chem. Solids \textbf{1}, 209
(1957).

\bibitem{nonlinear2} V. E. Zaharov, V. S. L'vov, and S. S. Starobinets, Sov.
Phys. Usp. \textbf{17}, 896 (1975).

%self-energy

\bibitem{Abrikosov} A. A. Abrikosov, L. P. Gorkov, and I. E. Dzyaloshinski, 
\textit{Methods of Quantum Field Theory in Statistical Physics} (Prentice
Hall, Englewood Cliffs, N. J., 1963).

\bibitem{Fetter} A. L. Fetter and J. D. Walecka, \textit{Quantum Theory of
	Many Particle Systems} (McGraw-Hill, New York, 1971).

\bibitem{Mahan} G. D. Mahan, \textit{Many Particle Physics} (Plenum, New
York, 1990).

%Gilbert damping

\bibitem{Gilbert_damping} T. L. Gilbert, IEEE Trans. Magn. \textbf{40}, 3443
(2004).

%input_output theory

\bibitem{input_output1} C. W. Gardiner and M. J. Collett, Phys. Rev. A 
\textbf{31}, 3761 (1985).

\bibitem{input_output2} A. A. Clerk, M. H. Devoret, S. M. Girvin, F.
Marquardt, and R. J. Schoelkopf, Rev. Mod. Phys. \textbf{82}, 1155 (2010).



%YIG damping

\bibitem{Haiming_PRL} J. L. Chen, C. P. Liu, T. Liu, Y. Xiao, K. Xia, G. E.
W. Bauer, M. Z. Wu, and H. M. Yu, Phys. Rev. Lett. \textbf{120}, 217202
(2018).

\bibitem{YIG_damping1} H. Chang, P. Li, W. Zhang, T. Liu, A. Hoffmann, L.
Deng, and M. Wu, IEEE Magn. Lett. \textbf{5}, 1 (2014).

\bibitem{surface_sensitivity} A. Aqeel, I. J. Vera-Marun, B. J. van Wees,
and T. T. M. Palstra, J. Appl. Phys. \textbf{116}, 153705 (2014).

\bibitem{Ando} T. Ando, A. B. Fowler, and F. Stern, Rev. Mod. Phys. \textbf{%
	54}, 437 (1982).

%Brillouin light scattering

\bibitem{Brillouin_light_scattering} G. Srinivasan, C. E. Patton, and P. R.
Emtage, J. Appl. Phys. \textbf{61}, 2318 (1987).

\bibitem{Nobel} P. A. Gr\"{u}nberg, Rev. Mod. Phys. \textbf{80}, 1531 (2008).

\bibitem{Hashimoto} Y. Hashimoto, S. Daimon, R. Iguchi, Y. Oikawa, K. Shen,
K. Sato, D. Bossini, Y. Tabuchi, T. Satoh, B. Hillebrands, G. E. W. Bauer,
T. H. Johansen, A. Kirilyuk, T. Rasing, and E. Saitoh, Nat. Comm. \textbf{8}%
, 15859 (2017).

%single site

\bibitem{Velicky} B. Velicky, S. Kirkpatrick, and H. Ehrenreich, Phys. Rev. 
\textbf{175}, 747 (1968).

\bibitem{exchange_1969} R. E. De Wames and T. Wolfram, Appl. Phys. Lett. 
\textbf{15}, 297 (1969).

\bibitem{exchange_six_order} T. Wolfram and R. E. De Wames, Phys. Rev. Lett. 
\textbf{24}, 1489 (1970).



%transport

\bibitem{Haug} H. Haug and A. P. Jauho, \textit{Quantum Kinetics in
	Transport and Optics of Semiconductors} (Springer, Berlin, 1996).

\bibitem{cornelissen} L. J. Cornelissen, K. J. H. Peters, G. E. W. Bauer, R.
A. Duine, and B. J. van Wees, Phys. Rev. B \textbf{94}, 014412 (2016).

\bibitem{Boltzmann} B. Flebus, K. Shen, T. Kikkawa, K. Uchida, Z. Qiu, E.
Saitoh, R. A. Duine, and G. E. W. Bauer, Phys. Rev. B \textbf{95}, 144420
(2017). %cavity spintronics

\bibitem{first_2010} O. O. Soykal, and M. E. Flatt\'e, Phy. Rev. Lett. 
\textbf{104}, 077202 (2010).

\bibitem{second} H. Huebl, C. W. Zollitsch, J. Lotze, F. Hocke, M.
Greifenstein, A. Marx, R. Gross, and S. T. B. Goennenwein, Phy. Rev. Lett. 
\textbf{111}, 127003 (2013).

\bibitem{third} Y. Tabuchi, S. Ishino, T. Ishikawa, R. Yamazaki, K. Usami,
and Y. Nakamura, Phy. Rev. Lett. \textbf{113}, 083603 (2014).

\bibitem{fourth} X. Zhang, C.-L. Zou, L. Jiang, and H. X. Tang, Phy. Rev.
Lett. \textbf{113}, 156401 (2014).

%dynamical surface roughness

\bibitem{dynamical1} M. Weiler, L. Dreher, C. Heeg, H. Huebl, R. Gross, M.
S. Brandt, and S. T. B. Goennenwein, Phys. Rev. Lett \textbf{106}, 117601
(2011).

\bibitem{dynamical2} R. Sasaki, Y. Nii, Y. Iguchi, and Y. Onose, Phys. Rev.
B \textbf{95}, 020407(R) (2017).

\bibitem{dynamical3} R. Verba, I. Lisenkov, I. Krivorotov, V. Tiberkevich,
and A. Slavin, Phys. Rev. Appl. \textbf{9}, 064014 (2018).

\end{thebibliography}
\end{document}